\documentclass[twocolumn,aps,prb,10pt,superscriptaddress,longbibliography,floatfix]{revtex4-2}
\usepackage{amsmath}
\usepackage{bm}
\usepackage{glossaries}
\usepackage{graphicx}
\usepackage{multirow}
\usepackage{textcomp}
\usepackage{threeparttable}
\usepackage{needspace}
\usepackage[usenames,dvipsnames,svgnames]{xcolor}
\usepackage[normalem]{ulem}
\usepackage[english]{babel}
\usepackage{hyperref}
\addto\extrasenglish{}
\hypersetup{
pdfnewwindow=true,
colorlinks=true,
linkcolor=Blue,
citecolor=Blue,
filecolor=Blue,
urlcolor=Blue
}
\usepackage{siunitx}
\newacronym{dft}{DFT}{density functional theory}
\newacronym{nep}{NEP}{neuroevolution potential}
\newacronym{mlp}{MLP}{machine-learned potential}
\newacronym{rmse}{RMSE}{root-mean-square error} 
\newacronym{zbl}{ZBL}{Ziegler--Biersack--Littmark}
\newacronym{hnemd}{HNEMD}{homogeneous non-equilibrium molecular dynamics}
\newacronym{nemd}{NEMD}{non-equilibrium molecular dynamics}
\DeclareSIUnit\angstrom{\text{Å}}
\DeclareSIUnit\atom{\text{atom}}

\begin{document}

\title{A Transferable Machine-Learned Potential Enabling Large-Scale Copper--Carbon Simulations from Equilibrium to Extreme Conditions}

\author{Feiyu Zhou}
\thanks{These authors contributed equally to this work.}
\affiliation{
Laboratory for multiscale mechanics and medical science, SV LAB, School of Aerospace, Xi'an Jiaotong University, Xi'an, Shaanxi, 710049, China
}

\author{Wei Qiu}
\thanks{These authors contributed equally to this work.}
\affiliation{Department of Physical Chemistry, School of Chemistry, The Raymond and Beverly Sackler Faculty of Exact Sciences and
The Sackler Center for Computational Molecular and Materials Science, Tel Aviv University, Tel Aviv 6997801, Israel}

\author{Yan Chen}
\affiliation{
Laboratory for multiscale mechanics and medical science, SV LAB, School of Aerospace, Xi'an Jiaotong University, Xi'an, Shaanxi, 710049, China
}

\author{Jiahui Liu}
\email{liujiahui@zgci.ac.cn}
\affiliation{Zhongguancun Institute of Artificial Intelligence, Beijing 100094, China}

\author{Yilun Liu}
\affiliation{
Laboratory for multiscale mechanics and medical science, SV LAB, School of Aerospace, Xi'an Jiaotong University, Xi'an, Shaanxi, 710049, China
}

\author{Shengying Yue}
\email{syyue@xjtu.edu.cn}
\affiliation{
Laboratory for multiscale mechanics and medical science, SV LAB, School of Aerospace, Xi'an Jiaotong University, Xi'an, Shaanxi, 710049, China
}

\author{Penghua Ying}
\email{penghua@xjtu.edu.cn}
\affiliation{
Laboratory for multiscale mechanics and medical science, SV LAB, School of Aerospace, Xi'an Jiaotong University, Xi'an, Shaanxi, 710049, China
}

\date{\today}

\begin{abstract}
\noindent{\textbf{\large Abstract}}
\noindent
Cu--C materials and interfaces are central to diamond thermal management, metallization, graphene growth and carbon-coating technologies, but atomistic simulations across these applications require a potential that describes metallic Cu, diverse carbon bonding, heterogeneous Cu--C interactions and close atomic encounters within a single framework. Here we develop NEP-CuC, a transferable neuroevolution potential trained on a train dataset of 2,727 uniformly relabelled configurations selected from a heterogeneous Cu--C candidate pool. Disagreement within an ensemble of five independently trained models provides an internal indicator of model sensitivity. A Ziegler–Biersack–Littmark (ZBL) correction is incorporated to describe short-range nuclear repulsion during close atomic encounters, while a runtime D3(BJ) correction supplements dispersion-sensitive calculations.  Tests on source-derived held-out and independent external datasets, together with structural, elastic, defect, interfacial, vibrational and finite-temperature transport benchmarks, establish broad but property-dependent physical fidelity across elemental Cu, diverse carbon environments and heterogeneous Cu--C configurations. Large-scale simulations of amorphous-carbon formation and orientation-dependent shock compression of Cu further test transferability under strongly non-equilibrium conditions. For Cu/diamond interfaces, the same potential describes deposition, subsequent annealing and tensile response: preparation history leaves persistent differences in interphase structure, while the mean peak nominal tensile stress at 500 K decreases from $12.35\pm0.71$ to $6.65\pm0.98$~GPa as the incident Cu energy during the initial deposition stage increases from 1 to 50 eV. Together, NEP-CuC provides an atomistic framework for linking processing, interfacial structure and physical response across Cu--C systems, with direct relevance to Cu/diamond thermal-management interfaces, diamond metallization and non-equilibrium Cu--C processing.
\end{abstract}
\maketitle

\noindent\textbf{INTRODUCTION}

\noindent The performance of Cu--C materials is governed by the properties of their constituents and the interfaces formed during processing. Cu/diamond composites combine the high thermal conductivity of diamond with the ductility and processability of Cu, while interfacial structure and chemistry control adhesion and heat transfer~\cite{Graebner1992Diamond,Jia2021CuDiamondReview,Wang2001CuDiamond,Liu2018CuDiamond}. Metallization, carbide-forming interlayers and organic monolayers are used to modify these interfaces~\cite{Zhang2018DiamondMetallization,Chang2018TiCuDiamond,xu2021weaker,Sun2025CuDiamond}. Related Cu--C environments occur during graphene growth on Cu~\cite{Li2009GrapheneCu,Yu2011GrapheneCu} and energetic carbon deposition~\cite{Caro2018TaCGrowth,Caro2020CarbonDeposition}. Such processes produce structures ranging from weakly bound interfaces to reconstructed and compositionally mixed regions. Resolving their evolution requires atomistic simulations over length and time scales beyond the routine reach of first-principles molecular dynamics~\cite{deringer2019machine}.

Classical interatomic potentials make large-scale molecular dynamics tractable, but a single Cu--C parameterization must accommodate markedly different bonding environments. Existing approaches include element-specific potentials coupled through Lennard--Jones cross interactions, second-nearest-neighbour modified embedded-atom method (2NN-MEAM) potentials~\cite{2NNMEAM2021Copper}, reactive force fields~\cite{Kowalik2021GrapheneCuReaxFF} and analytical bond-order models~\cite{Zhou2015CarbonBOP,NIST2015ZhouCCu}. Each has a defined domain of applicability. A transferable Cu--C description must capture metallic bonding in crystalline, defective, liquid and compressed Cu, alongside covalent bonding in $\mathrm{sp}$, $\mathrm{sp}^{2}$ and $\mathrm{sp}^{3}$ carbon environments. It must also describe heterogeneous interactions across weakly bound and reconstructed interfaces, together with short-range repulsion during energetic collisions. Maintaining accuracy across these distinct regimes is the central modelling challenge.

\Glspl{mlp} address this challenge by fitting flexible many-body representations to first-principles reference data~\cite{behler2007nn,deringer2019machine,Batzner2022NequIP,Chen2022M3GNet}. Recent Cu--C models have described carbon growth and deposition on solid Cu~\cite{Zhang2024CarbonGrowthCu,liu_generalized_2025}, graphene adsorption and growth on liquid Cu~\cite{Gao2022LiquidCuMTP,Rein2024LiquidCuGrowth}, hydrocarbon chemistry on Cu(111)~\cite{Xu2024CuCHNNP}, and Cu/diamond interfacial heat transport~\cite{Sun2025CuDiamond}. These studies establish the capabilities of models developed for specific structural or chemical environments. A complementary question is whether a single potential can describe the Cu--C environments outlined above with sufficient accuracy for their characteristic physical observables. For weakly bound interfaces, this description must also account for dispersion interactions beyond the local model cutoff.

Here we develop NEP-CuC, a transferable Cu--C potential based on the \gls{nep} framework~\cite{fan2021neuroevolution,Fan2022NEP}, and progressively assess its accuracy and transferability. Disagreement within an ensemble of five independently trained models provides an internal indicator of model sensitivity. Screened nuclear repulsion treats close atomic encounters~\cite{ziegler1985stopping}, while runtime D3(BJ) supplements the same trained parameters in dispersion-sensitive calculations~\cite{Ying2024NEPD3}. We first test configuration-level predictions, then equilibrium energy landscapes, lattice dynamics and finite-temperature phonon transport. Amorphous-carbon formation and Cu shock compression extend this assessment to complementary non-equilibrium elemental limits. Finally, Cu deposition on diamond combines elemental rearrangement with heterogeneous interface formation, linking processing history to interphase structure and subsequent mechanical response.

\vspace{0.5cm}
\noindent\textbf{RESULTS}

\noindent\textbf{Model development, prediction accuracy and computational performance}

\begin{figure*}[t]
    \centering
    \includegraphics[width=1.8\columnwidth]{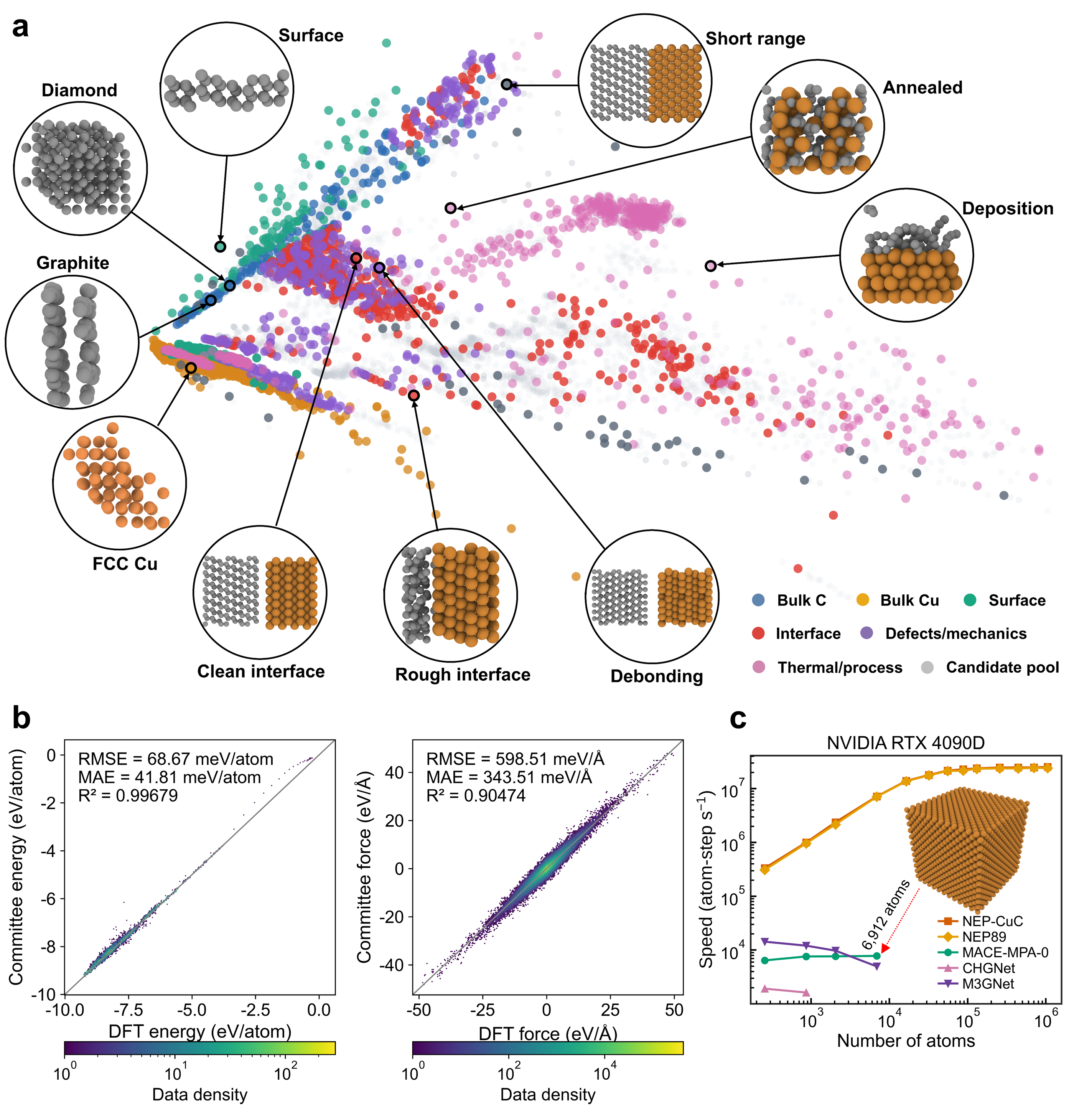}
    \caption{\textbf{Configuration selection, prediction errors and throughput of NEP-CuC.} \textbf{a}, Two-dimensional principal-component projection of the full 15,979-geometry candidate inventory (grey) and the final 2,727-configuration development set (coloured by structural class), using descriptors from NEP-CuC model M1. Insets show representative structures. Configuration selection was performed separately using NEP89 descriptors and farthest-point sampling. \textbf{b}, Ensemble-mean energy and force predictions from NEP-CuC models M1--M5 against density functional theory (DFT) references on the source-derived held-out benchmark. Colour indicates point density. \textbf{c}, Molecular dynamics throughput of NEP-CuC model M1 versus atom count on one NVIDIA RTX 4090D graphics processing unit (GPU). Comparison models are NEP89~\cite{Liang2026NEP89}, MACE-MPA-0~\cite{mace2025Batatia}, CHGNet~\cite{deng2023chgnet} and M3GNet~\cite{Chen2022M3GNet}. Implementation and timing details are specified in Methods and Supplementary Note~S8.}
    \label{fig:trainingRMSE}
\end{figure*}

\noindent The NEP-CuC development dataset comprises 2,727 unique configurations uniformly relabelled with energies, forces and virials using a common first-principles protocol. Publicly available Cu, C and Cu--C structural datasets were combined with purpose-built configurations to form a 15,979-geometry candidate inventory spanning crystals, defects, surfaces, interfaces, thermal disorder, deformation and close atomic encounters. Configurations were selected by structural-family stratification and farthest-point sampling using NEP89 descriptors~\cite{Liang2026NEP89}, followed by geometry-based de-duplication. \autoref{fig:trainingRMSE}(a) visualizes the candidate inventory and final development set using descriptors from the fitted NEP-CuC model. Details of dataset construction, configuration selection, \gls{dft} relabelling and model training/validation are provided in Methods and Supplementary Notes~S1, S2 and S3.

Predictive accuracy was evaluated for the five members of the NEP-CuC ensemble (M1--M5) and their ensemble-mean predictions. On 10,961 source-derived, label-held-out configurations~\cite{Sun2025CuDiamond,yang_tuning_2026,Song2024UNEPv1,kang_intelligent_2026,liu_enhanced_2026,Fan_2024}, the ensemble mean yields energy and force \glspl{rmse} of \SI{68.67}{\milli\electronvolt\per\atom} and \SI{598.51}{\milli\electronvolt\per\angstrom}, respectively (\autoref{fig:trainingRMSE}b). The corresponding stress \gls{rmse} is \SI{3.031}{\giga\pascal}. Because the source libraries also contributed geometries to the candidate pool, these errors characterize generalization within represented structural domains. Relative to the mean individual-model RMSE, ensemble averaging reduces energy, force and stress errors by 4.3\%, 8.5\% and 15.0\%, respectively (Supplementary Fig.~S1). Source-resolved errors vary substantially across structural classes (Supplementary Fig.~S2).

Transferability was further evaluated on two external datasets excluded from all stages of model development. For 5,417 configurations describing carbon deposition and growth on Cu(111)~\cite{liu_generalized_2025}, NEP-CuC yields energy and force \glspl{rmse} of \SI{122.99}{\milli\electronvolt\per\atom} and \SI{666.88}{\milli\electronvolt\per\angstrom}, respectively (Supplementary Fig.~S3a). For context, Liu \textit{et al.} reported corresponding test \glspl{rmse} of \SI{57.752}{\milli\electronvolt\per\atom} and \SI{555.817}{\milli\electronvolt\per\angstrom} for their deposition-focused \gls{nep}. This comparison is not strictly one-to-one because the two models were evaluated on different configuration sets, and the source study's reference calculations employ the Perdew--Burke--Ernzerhof (PBE) functional with DFT-D3 dispersion corrections whereas NEP-CuC was fitted to bare-PBE labels. The present errors therefore incorporate both out-of-domain transfer and reference-protocol differences. Nevertheless, the force error remains close to that of the deposition-focused model despite the complete exclusion of these configurations from NEP-CuC development.

On the 450-configuration external carbon test set~\cite{PhysRevB.111.094205}, NEP-CuC gives energy and force \glspl{rmse} of \SI{59.51}{\milli\electronvolt\per\atom} and \SI{1.034}{\electronvolt\per\angstrom}, respectively (Supplementary Fig.~S3b). On the same benchmark, Wang \textit{et al.} reported \SI{45}{\milli\electronvolt\per\atom} and \SI{0.732}{\electronvolt\per\angstrom} for their carbon-specific \gls{nep}, while the Gaussian approximation potential (GAP) models GAP17 and GAP22 yielded force \glspl{rmse} of \SI{1.14}{\electronvolt\per\angstrom} and \SI{1.04}{\electronvolt\per\angstrom}, respectively. The source study further showed that the larger aggregate force error partly originates from liquid configurations with large atomic forces. Despite no exposure to this dataset during training, NEP-CuC therefore achieves a force \gls{rmse} comparable to those of GAP17 and GAP22.

The compact \gls{nep} representation provides the throughput required for the large-scale simulations considered below. In end-to-end microcanonical (NVE) benchmarks of periodic face-centred cubic (fcc) Cu, a single NEP-CuC model reaches approximately $2.4\times10^{7}$ atom steps~s$^{-1}$ on a single NVIDIA GeForce RTX 4090D graphics processing unit (GPU) for systems containing 131,072--1,048,576 atoms (\autoref{fig:trainingRMSE}c, Supplementary Note~S8), with throughput comparable to NEP89~\cite{Liang2026NEP89}. Under the same benchmark protocol, MACE-MPA-0~\cite{mace2025Batatia} and M3GNet~\cite{Chen2022M3GNet} were evaluated for systems up to 6,912 atoms and CHGNet~\cite{deng2023chgnet} up to 864 atoms. Together with the GPU-accelerated implementation in Graphics Processing Units Molecular Dynamics (\textsc{gpumd})~\cite{zheyong2025GPUMD}, this efficiency enables the large-scale non-equilibrium simulations below, including million-atom shock compression and deposition-generated Cu/diamond assemblies.

\vspace{0.5cm}
\noindent\textbf{Equilibrium properties, energy profiles and lattice dynamics}

\noindent Equilibrium properties test the physical accuracy of NEP-CuC beyond aggregate prediction errors. \autoref{tab:static_properties_cu_c} compares NEP-CuC predictions with experimental and \gls{dft} reference data and predictions from the Mishin embedded-atom method (EAM) potential~\cite{mishin2001structural}, Tersoff~\cite{tersoff1989modeling}, 2NN-MEAM~\cite{2NNMEAM2021Copper}, MACE-MPA-0~\cite{mace2025Batatia} and NEP89~\cite{Liang2026NEP89}. Property definitions and the reference states used for energy calculations are detailed in Methods and Supplementary Note~S4.

\begin{table*}[t!]
\caption{\textbf{Static properties of Cu, carbon and Cu--C environments.} NEP-CuC predictions are compared with Mishin EAM~\cite{mishin2001structural}, Tersoff~\cite{tersoff1989modeling}, 2NN-MEAM~\cite{2NNMEAM2021Copper}, MACE-MPA-0~\cite{mace2025Batatia} and NEP89~\cite{Liang2026NEP89}. Reference data include experimental measurements, literature density functional theory (DFT) results and present DFT calculations. NEP-CuC values are ensemble means $\pm$ standard error of the mean (SEM) across models M1--M5 ($n=5$).}
\label{tab:static_properties_cu_c}
\centering
\begingroup
\setlength{\tabcolsep}{2.5pt}
\resizebox{\textwidth}{!}{%
\begin{tabular}{llllllllll}
\hline\hline
Property & Mishin EAM & Tersoff & 2NN-MEAM & MACE-MPA-0 & NEP89 & Exp. & Literature DFT & Present DFT & NEP-CuC \\
\hline
\multicolumn{10}{l}{\textbf{Elemental Cu}} \\
$a_0^{\rm Cu}$ & 3.615 &  & 3.615\cite{2NNMEAM2021Copper} & 3.640 & 3.557 & 3.615\cite{Ma2021PointDefects} & 3.634\cite{Ma2021PointDefects} & 3.635 & 3.6335 $\pm$ 0.0004 \\
$C_{11}^{\rm Cu}$ & 170.5 &  & 188.4\cite{2NNMEAM2021Copper} & 157.0 & 209.1 & 169.1\cite{Ledbetter1974Copper} & 178.484\cite{Ma2021PointDefects} & 182.3 & 173.0 $\pm$ 2.4 \\
$C_{12}^{\rm Cu}$ & 124.7 &  & 118.9\cite{2NNMEAM2021Copper} & 117.3 & 149.0 & 122.2\cite{Ledbetter1974Copper} & 129.262\cite{Ma2021PointDefects} & 117.3 & 125.3 $\pm$ 1.2 \\
$C_{44}^{\rm Cu}$ & 76.3 &  & 72.2\cite{2NNMEAM2021Copper} & 78.5 & 97.8 & 75.42\cite{Ledbetter1974Copper} & 74.128\cite{Ma2021PointDefects} & 74.2 & 77.3 $\pm$ 0.7 \\
$E_{\rm f}^{\rm vac,Cu}$ & 1.275 &  & 2.265 & 1.067 & 1.267 & 1.19\cite{Hehenkamp1992CuVacancy} & 1.064\cite{Ma2021PointDefects} & 1.079 & 1.074 $\pm$ 0.013 \\
$\gamma_{\rm surf}^{\rm Cu(100)}$ & 1.346 &  & 2.407 & 1.418 & 2.010 &  & 1.48\cite{Patra2017Surfaces} & 1.412 & 1.461 $\pm$ 0.005 \\
$\gamma_{\rm surf}^{\rm Cu(110)}$ & 1.476 &  & 2.477 & 1.514 & 2.105 &  & 1.63\cite{Patra2017Surfaces} & 1.519 & 1.532 $\pm$ 0.007 \\
$\gamma_{\rm surf}^{\rm Cu(111)}$ & 1.240 &  & 2.043 & 1.228 & 1.849 &  & 1.33\cite{Patra2017Surfaces} & 1.257 & 1.294 $\pm$ 0.007 \\
$\gamma_{\rm ISF}^{\rm Cu}$ & 45 &  & 64 & 28 & 16 & 45\cite{mishin2001structural} & 41\cite{Branicio2013CuISF} &  & 44 $\pm$ 11 \\
$\gamma_{\rm USF}^{\rm Cu}$ & 162 &  & 367 & 155 & 169 &  & 163.4\cite{Zhang2014CuFault} & 176 & 173 $\pm$ 7 \\
$E_{\rm f}^{\rm SIA,Cu,\langle100\rangle\mathrm{db}}$ &  &  &  &  &  & 3.47\cite{Ma2021PointDefects} & 3.088\cite{Ma2021PointDefects} &  & 3.152 $\pm$ 0.013 \\
\hline
\multicolumn{10}{l}{\textbf{Elemental C}} \\
$a_0^{\rm dia}$ &  & 3.566 & 3.567\cite{2NNMEAM2021Copper} & 3.578 & 3.565 & 3.5671\cite{Yamanaka1994Influence} & 3.5569\cite{Clerc2005Diamond} & 3.572 & 3.573 $\pm$ 0.002 \\
$C_{11}^{\rm dia}$ &  & 1073.3 & 1091.9\cite{2NNMEAM2021Copper} & 741.6 & 908.3 & 1080.4\cite{Vogelgesang1996Brillouin} & 1042\cite{Qamar2023ACE} & 1053.7 & 982.8 $\pm$ 33.2 \\
$C_{12}^{\rm dia}$ &  & 102.7 & 120.4\cite{2NNMEAM2021Copper} & 175.0 & 174.5 & 127\cite{Vogelgesang1996Brillouin} & 131\cite{Qamar2023ACE} & 129.7 & 216.4 $\pm$ 14.5 \\
$C_{44}^{\rm dia}$ &  & 641.1 & 653.0\cite{2NNMEAM2021Copper} & 505.5 & 622.7 & 576.6\cite{Vogelgesang1996Brillouin} & 556\cite{Qamar2023ACE} & 559.0 & 532.0 $\pm$ 21.9 \\
$E_{\rm f}^{\rm vac,dia}$ &  & 3.425 & 8.710 & 4.497 & 4.345 &  & 6.76\cite{Fatomi2022DiamondVacancy} & 6.604 & 3.726 $\pm$ 0.174 \\
$E_{\rm f}^{\rm split\langle100\rangle,dia}$ &  & 10.065 &  & 7.634 & 8.267 &  & 11.01\cite{Kuganathan2023DiamondDefects} & 11.217 & 7.789 $\pm$ 0.176 \\
$\gamma_{\rm surf}^{\rm dia(110)-(1\times1)}$ &  & 4.064 & 1.322 & 3.015 & 4.343 &  & 5.76\cite{Qamar2023ACE} & 5.178 & 4.237 $\pm$ 0.150 \\
$\gamma_{\rm surf}^{\rm dia(111)-3DB}$ &  & 14.077 & 21.997 & 11.172 & 11.905 &  & 12.93\cite{Qamar2023ACE} & 12.292 & 12.438 $\pm$ 0.089 \\
$d_0^{\rm graphite}$ &  &  &  & 3.972 & 4.153 & 3.354\cite{Baskin1955Graphite} & 3.356\cite{Chen2013Graphite} & 3.369 & 3.997 $\pm$ 0.343 \\
$\Delta E_{\rm graphite-dia}$ &  &  & -0.260\cite{2NNMEAM2021Copper} & -0.147 & -0.066 & -0.022\cite{White2021GraphiteDiamond} & -0.029\cite{Qamar2023ACE} & -0.081 & -0.1227 $\pm$ 0.0102 \\
$a_0^{\rm gr}$ &  & 2.530 & 2.470\cite{2NNMEAM2021Copper} & 2.451 & 2.477 & 2.459\cite{Baskin1955Graphite} & 2.463\cite{Qian2021CarbonPotentials} & 2.468 & 2.471 $\pm$ 0.001 \\
$Y_{\rm 2D}^{\rm gr}$ &  & 407.591 & 281.168 & 283.520 & 307.295 & 340\cite{Lee2008Graphene} & 345\cite{Kudin2001Nanoshell} &  & 286.492 $\pm$ 11.271 \\
$E_{\rm f}^{\rm vac,gr}$ &  & 6.890 & 9.713 & 6.802 & 5.605 &  & 7.98\cite{Rodrigo2016GrapheneVacancy} &  & 5.878 $\pm$ 0.108 \\
$\kappa_{\rm bend}^{\rm gr}$ &  & 1.038 & 0.235 & 1.757 & 2.078 &  & 1.44\cite{Wei2013GrapheneBending} &  & 1.296 $\pm$ 0.083 \\
$\Delta E_{\rm curv}^{(10,10)}$ &  & 0.029 & 0.007 & 0.050 & 0.059 &  & 0.045\cite{Kurti1998CNT} &  & 0.037 $\pm$ 0.002 \\
$\Delta E_{\rm curv}^{(17,0)}$ &  & 0.031 & 0.007 & 0.051 & 0.061 &  & 0.045\cite{Kurti1998CNT} &  & 0.038 $\pm$ 0.002 \\
$C_{\rm ax}^{(10,10)}$ &  & 1100.848 & 761.093 & 744.993 & 790.193 &  & 836.273\cite{SanchezPortal1999CNT} &  & 761.039 $\pm$ 28.611 \\
\hline
\multicolumn{10}{l}{\textbf{Binary Cu--C}} \\
$E_{\rm sol}^{\rm C@oct}$ &  &  & 6.460 & 2.611 & 2.162 &  & 1.57\cite{Zhou2015CarbonCu} & 2.671 & 2.473 $\pm$ 0.064 \\
$E_{\rm sol}^{\rm C@tet}$ &  &  & 3.504 & 3.871 & 3.641 &  & 2.73\cite{Zhou2015CarbonCu} & 4.051 & 3.764 $\pm$ 0.117 \\
$E_{\rm sol}^{\rm C@sub}$ &  &  & 8.040 & 4.395 & 4.342 &  & 3.14\cite{Zhou2015CarbonCu} & 4.288 & 4.455 $\pm$ 0.094 \\
$d_{\rm eq}^{\rm Cu/dia(100)}$ &  &  & 1.631 & 1.725 & 1.673 &  & 1.549\cite{Yang2021CuDiamond} & 1.624 & 1.633 $\pm$ 0.007 \\
$W_{\rm ad}^{\rm Cu/dia(100)}$ &  &  & 3.449 & 1.887 & 4.685 &  & 4.204\cite{Yang2021CuDiamond} & 3.707 & 3.253 $\pm$ 0.082 \\
\hline\hline
\end{tabular}%
}
\vspace{0.4ex}
\scriptsize
\textit{Notes.} Abbreviations are intrinsic stacking fault (ISF), unstable stacking fault (USF), self-interstitial atom (SIA), dumbbell (db), three dangling bonds (3DB), diamond (dia), graphene (gr), octahedral (oct), tetrahedral (tet) and substitutional (sub). Lengths are in \AA{}, elastic constants in GPa, and defect and solution energies in eV. Surface and adhesion energies are in $\mathrm{J\,m^{-2}}$, and stacking-fault energies in $\mathrm{mJ\,m^{-2}}$. Relative and curvature energies are in $\mathrm{eV\,atom^{-1}}$, graphene moduli in $\mathrm{N\,m^{-1}}$, bending rigidities in eV, and nanotube axial stiffnesses in $\mathrm{eV\,\text{\AA}^{-1}}$. Blank cells indicate unavailable, inapplicable or uncalculated values. For $d_0^{\rm graphite}$ and $\Delta E_{\rm graphite-dia}$, NEP-CuC predictions include D3(BJ)~\cite{Ying2024NEPD3}, and present DFT values use PBE-D3(BJ). Other NEP-CuC and present DFT entries use bare NEP-CuC and PBE, respectively. Calculation protocols and reference-state conventions are given in Methods and Supplementary Note~S4.
\endgroup
\end{table*}

For elemental Cu, NEP-CuC closely reproduces present \gls{dft} values of the equilibrium lattice parameter and vacancy formation energy, while the three independent elastic constants differ from the present \gls{dft} values by approximately $4$--$7\%$ (\autoref{tab:static_properties_cu_c}). The low-index surface energies retain the ordering $\gamma_{111}<\gamma_{100}<\gamma_{110}$. Along the $\{111\}\langle112\rangle$ slip pathway, the unstable stacking-fault energy is $173$ versus $176$~mJ\,m$^{-2}$ from DFT. The intrinsic stacking-fault energy is $44$~mJ\,m$^{-2}$, close to the experimental estimate of $45$~mJ\,m$^{-2}$. These properties provide the equilibrium benchmarks for the shock and plasticity simulations below.

The Cu--C benchmarks probe cross-species interactions through dilute C incorporation and a coherent Cu/diamond(100) interface. NEP-CuC retains the \gls{dft} ordering of octahedral, tetrahedral and substitutional C solution energies, with absolute errors of $0.20$, $0.29$ and $0.17$~eV, respectively. At the coherent interface, the mean equilibrium separation is $1.633$~\AA{} versus $1.624$~\AA{} from DFT, whereas the mean work of adhesion is underestimated by approximately $12\%$ ($3.253$ versus $3.707$~J\,m$^{-2}$). The equilibrium separation is thus reproduced more closely than the adhesion energy in relative terms.

The carbon benchmarks assess performance across distinct bonding environments. NEP-CuC reproduces the diamond lattice parameter within $0.001$~\AA{} and predicts $C_{11}$ and $C_{44}$ within approximately $7\%$ and $5\%$ of the present DFT values. The graphene lattice parameter differs from DFT by $0.003$~\AA{}. Its bending rigidity, along with nanotube curvature energies and axial stiffness, is close to the available first-principles benchmarks (\autoref{tab:static_properties_cu_c}). The three-dangling-bond diamond(111) surface energy is also reproduced closely. Major discrepancies occur for diamond $C_{12}$ and localized defects: diamond vacancy and split-$\langle100\rangle$ interstitial formation energies are underestimated by approximately $44\%$ and $31\%$, respectively.

\begin{figure}[ht]
    \centering
    \includegraphics[width=1\columnwidth]{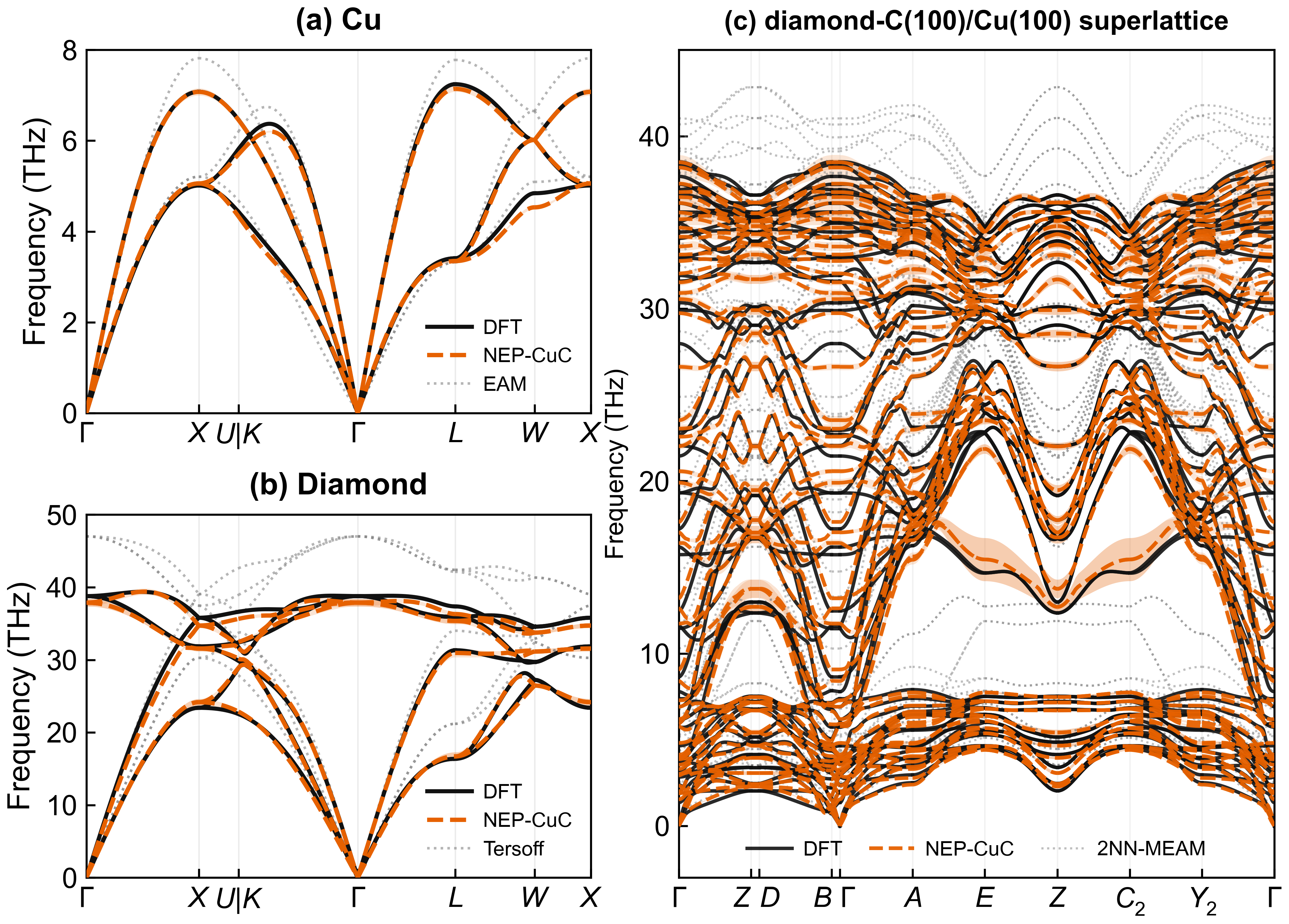}
    \caption{\textbf{Phonon dispersions of Cu, diamond and a coherent Cu/diamond superlattice.} \textbf{a--c}, fcc Cu, diamond and C(100)/Cu(100) superlattice, respectively. Black curves show density functional theory (DFT), orange curves the NEP-CuC ensemble mean and grey curves the comparison potentials: Mishin EAM for Cu~\cite{mishin2001structural}, Tersoff for diamond~\cite{tersoff1989modeling} and 2NN-MEAM for the superlattice~\cite{2NNMEAM2021Copper}. Shading denotes the NEP-CuC ensemble mean $\pm$ SEM across models M1--M5 ($n=5$).}
    \label{fig:VacmHn_results}
\end{figure}

Energy profiles test the potential-energy landscape beyond values at individual minima (Supplementary Figs.~S4--S8). NEP-CuC closely reproduces the \gls{dft} energy--volume curves of Cu and diamond and shows good agreement with their small-strain energy responses over $\pm1\%$. It also reproduces the shape and unstable barrier of the Cu stacking-fault energy profile, while predicting a lower intrinsic stacking-fault energy than the DFT path reference. Along the constrained Cu/diamond separation path, NEP-CuC closely matches the \gls{dft} relative-energy curve, complementing the independently relaxed adhesion benchmark. With the \gls{zbl} correction included, frozen C--C, C--Cu and Cu--Cu pairs exhibit steep short-range repulsion. The C--Cu and Cu--Cu energies exceed the DFT references at the shortest separations.

Phonon dispersions further assess the local curvature of the potential-energy surface around the relaxed structures, connecting these energetic benchmarks to vibrational behaviour. \autoref{fig:VacmHn_results} compares NEP-CuC with \gls{dft} and the corresponding empirical potentials for independently relaxed structures (Methods and Supplementary Note~S4). These are Mishin EAM~\cite{mishin2001structural} for Cu, Tersoff~\cite{tersoff1989modeling} for diamond and 2NN-MEAM~\cite{2NNMEAM2021Copper} for the C(100)/Cu(100) superlattice. NEP-CuC closely follows the \gls{dft} branches for fcc Cu and generally improves on Tersoff for diamond acoustic branches and optical frequencies. For the superlattice, it captures the main features of the dense \gls{dft} spectrum, whereas 2NN-MEAM shows larger frequency shifts. These harmonic benchmarks motivate the following assessment of finite-temperature phonon transport.

\vspace{0.5cm}
\noindent\textbf{Finite-temperature phonon transport from bulk crystals to Cu/diamond interfaces}

\begin{figure}[t]
    \centering
    \includegraphics[width=1\columnwidth]{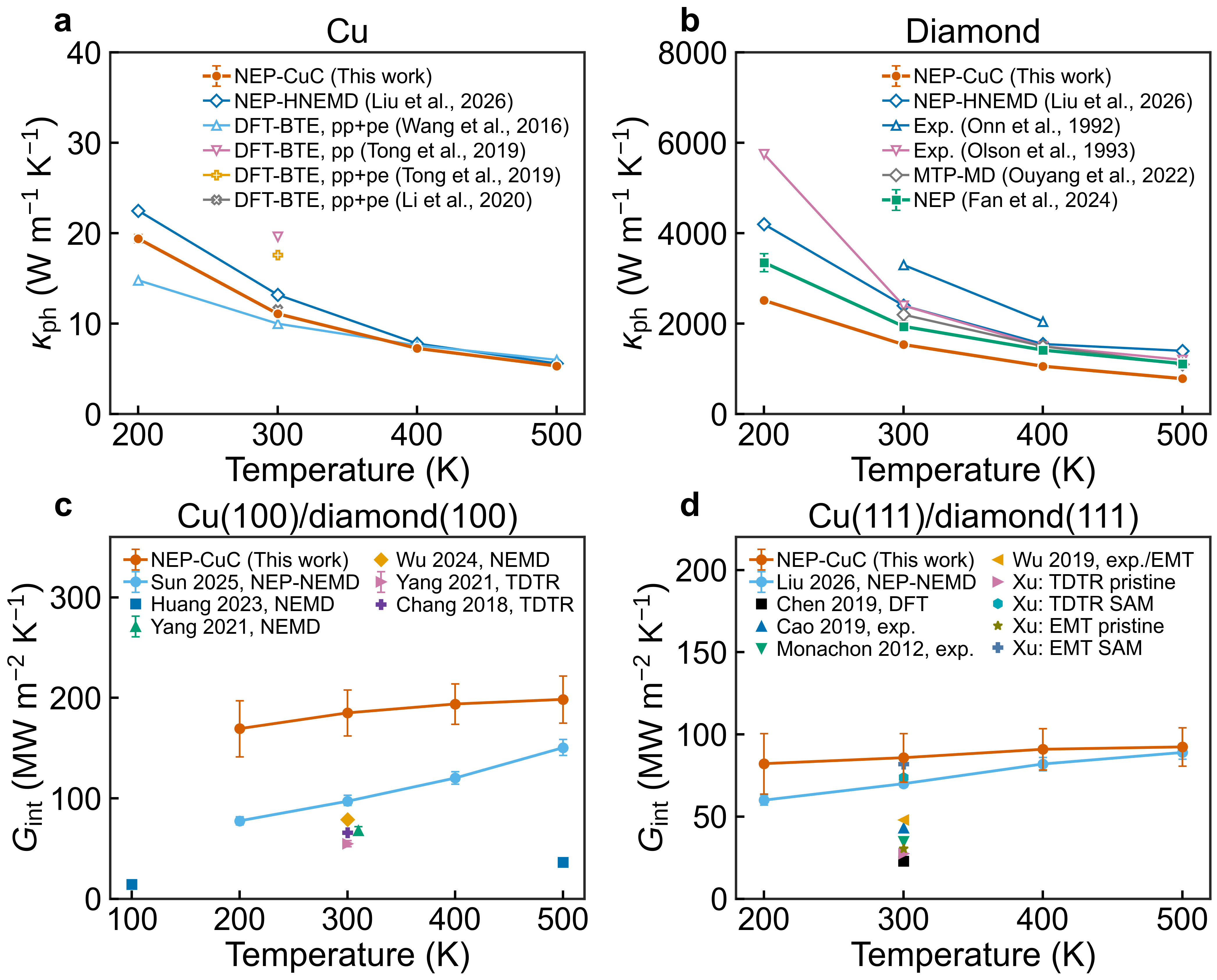}
    \caption{\textbf{Phonon transport in bulk crystals and Cu/diamond interfaces.}
    \textbf{a}, Lattice thermal conductivity of Cu calculated using homogeneous non-equilibrium molecular dynamics (HNEMD), compared with literature data from Refs.~\cite{liu_enhanced_2026,Wang2016MetalLTC,Tong2019MetalLTC,Li2020MetalTransport}.
    \textbf{b}, Lattice thermal conductivity of diamond calculated using HNEMD, compared with literature data from Refs.~\cite{liu_enhanced_2026,Ouyang2022DiamondMTP,Onn1992DiamondTC,Olson1993DiamondTC}. The carbon-specific NEP potential developed by Fan \textit{et al.}~\cite{Fan_2024} is also included.
    \textbf{c}, Phonon-mediated interfacial conductance of Cu(100)/diamond(100) calculated using non-equilibrium molecular dynamics (NEMD), compared with literature data from Refs.~\cite{Sun2025CuDiamond,Huang2023CuDiamondTransport,Yang2021CuDiamond,Wu2024CuDiamondVacancy,Chang2018TiCuDiamond}.
    \textbf{d}, Phonon-mediated interfacial conductance of Cu(111)/diamond(111) calculated using NEMD, compared with literature data from Refs.~\cite{liu_enhanced_2026,Chen2019CuDiamondTransport,Cao2019GrapheneInterlayer,Monachon2012MetalDiamond,Wu2019ElectroplatedCuDiamond,Xu2021ScalableSAM}.
    For NEP-CuC, symbols and error bars denote ensemble means and SEMs, respectively, across models M1--M5. In the legends, pp and pe denote phonon--phonon and phonon--electron scattering, respectively. BTE, MTP and MD denote the Boltzmann transport equation, moment tensor potential and molecular dynamics, respectively. TDTR denotes time-domain thermoreflectance, SAM denotes self-assembled monolayer, and EMT denotes effective-medium theory.}
    \label{fig:bench}
\end{figure}

\noindent Having assessed harmonic lattice dynamics, we next tested finite-temperature phonon transport in bulk Cu, diamond and coherent Cu/diamond interfaces (\autoref{fig:bench}). These observables probe collective dynamics beyond the harmonic force constants. Simulation protocols and statistical definitions are provided in Methods. For fcc Cu, using the \gls{hnemd} approach~\cite{Fan2019HNEMD}, NEP-CuC predicts a lattice thermal conductivity decreasing from $19.41\pm0.47$~W\,m$^{-1}$\,K$^{-1}$ at 200~K to $5.33\pm0.21$~W\,m$^{-1}$\,K$^{-1}$ at 500~K. The predicted temperature dependence and magnitude are consistent with the previous \gls{hnemd} results~\cite{liu_enhanced_2026} and with \gls{dft}-based Boltzmann-transport calculations for the phononic contribution~\cite{Wang2016MetalLTC,Tong2019MetalLTC,Li2020MetalTransport} (\autoref{fig:bench}a). These values describe lattice-mediated heat transport only and therefore exclude the dominant electronic contribution to the total thermal conductivity of metallic Cu.

Diamond presents a more demanding transport benchmark. NEP-CuC predicts $\kappa_{\mathrm{ph}}$ to decrease from $2514\pm50$~W\,m$^{-1}$\,K$^{-1}$ at 200~K to $784\pm24$~W\,m$^{-1}$\,K$^{-1}$ at 500~K (\autoref{fig:bench}b), with a value of $1538\pm64$~W\,m$^{-1}$\,K$^{-1}$ at 300~K. The model reproduces the characteristic decrease in thermal conductivity with increasing temperature, but the absolute values remain below the experimental and atomistic references shown in \autoref{fig:bench}b. Across 200--500~K, the predicted conductivities are also approximately 20--30\% lower than those obtained with the carbon-specific NEP potential developed by Fan \textit{et al.}~\cite{Fan_2024}. Thus, NEP-CuC captures the temperature dependence of diamond heat transport while retaining a systematic underestimation of its absolute conductivity. Force errors may contribute to this underestimation, as they can introduce additional phonon scattering in molecular dynamics simulations with \glspl{mlp}~\cite{Wu2024ForceError}.

We finally evaluated phonon-mediated heat transfer across coherent Cu/diamond interfaces using the \gls{nemd} approach (\autoref{fig:bench}c,d). Both orientations show only a moderate increase in ensemble-mean conductance with temperature, but a pronounced orientation dependence persists throughout the investigated range. For Cu(100)/diamond(100), the ensemble-mean conductance increases from $169\pm28$ to $198\pm23$~MW\,m$^{-2}$\,K$^{-1}$ between 200 and 500~K, whereas that of Cu(111)/diamond(111) increases from $82\pm18$ to $92\pm12$~MW\,m$^{-2}$\,K$^{-1}$. All five ensemble members predict $G_{100}>G_{111}$ at every temperature, with the ratio of the ensemble means remaining between 2.06 and 2.16. The ordering $G_{100}>G_{111}$ is therefore robust across the ensemble, despite variation in the absolute conductance.

The absolute interfacial conductance values exhibit appreciable model dependence. The inter-model standard deviation (SD) corresponds to 23--37\% of the ensemble mean for Cu(100)/diamond(100) and 28--50\% for Cu(111)/diamond(111). The predicted (111) conductance approaches previous \gls{nep}-based \gls{nemd} values~\cite{liu_enhanced_2026} at elevated temperatures, whereas the coherent (100) interface yields values above most of the published estimates shown in \autoref{fig:bench}c. For both orientations, the calculated conductances exceed several of the experimental estimates shown in \autoref{fig:bench}c,d. This difference may partly reflect the contrast between the ideal coherent interfaces considered here and experimentally prepared interfaces, where incomplete contact, defects and interfacial layers can alter heat transfer~\cite{Hopkins2010AlSiRoughness,wang2026interface}. In addition to the interfacial structures, the present phonon-only simulations omit electron--phonon coupling within Cu, which can modify thermal resistance and phonon energy redistribution~\cite{Majumdar2004ElectronPhonon,Wu2015ElectronPhonon,Lu2026GrapheneInterlayer}. Future simulations combining NEP-CuC with two-temperature NEMD are needed to assess the contribution of this coupling to the discrepancies with experiment. Because these literature values span different orientations, interface structures and preparation protocols, they provide a comparison of the conductance scale rather than a strictly matched quantitative benchmark.

\vspace{0.5cm}
\noindent\textbf{Moir\'e corrugation at graphene/Cu(111)}

\begin{figure}[t]
    \centering
    \includegraphics[width=\columnwidth]{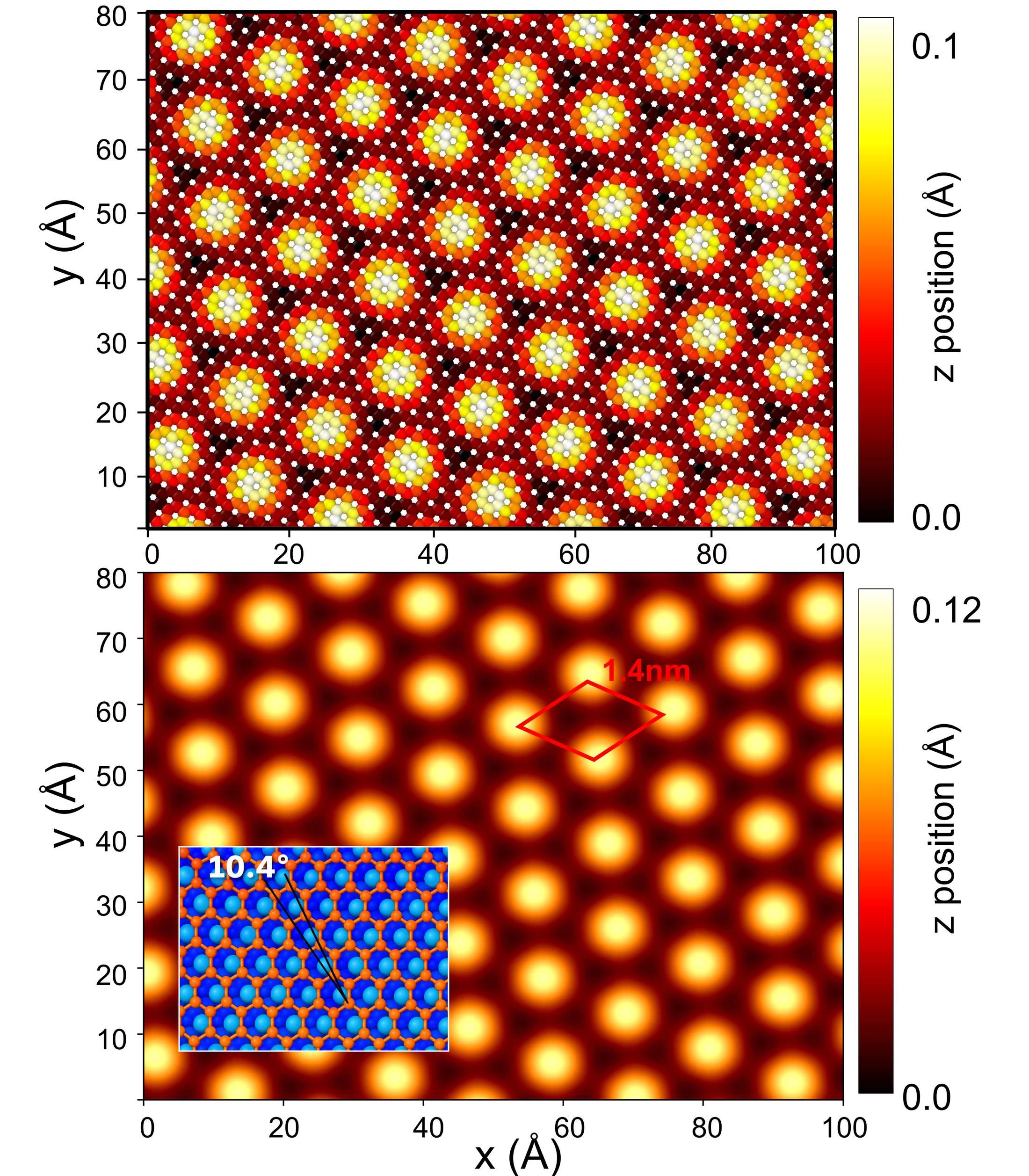}
    \caption{\textbf{Moir\'e corrugation of graphene on Cu(111).} Atom-resolved graphene heights (top) and the corresponding height field (bottom) after relaxation using NEP-CuC model M1 with runtime D3(BJ). D3(BJ) denotes the D3 dispersion correction with Becke--Johnson damping. Colour denotes out-of-plane displacement in \AA{}. The marked repeat corresponds to the commensurate cell. The morphology is qualitatively consistent with R10 graphene/Cu(111) superstructures observed by scanning tunnelling microscopy~\cite{Sule2014GrapheneCu}.}
    \label{fig:moire}
\end{figure}

\noindent Graphene/Cu(111) complements the coherent Cu/diamond benchmarks by testing a weakly bound, dispersion-sensitive interface. As in the graphite benchmarks in \autoref{tab:static_properties_cu_c}, runtime D3(BJ) supplements the local potential trained on bare-PBE labels. The additive correction accounts for dispersion without modifying the trained \gls{nep} parameters.

\begin{figure*}[ht]
    \centering
    \includegraphics[width=1.5\columnwidth]{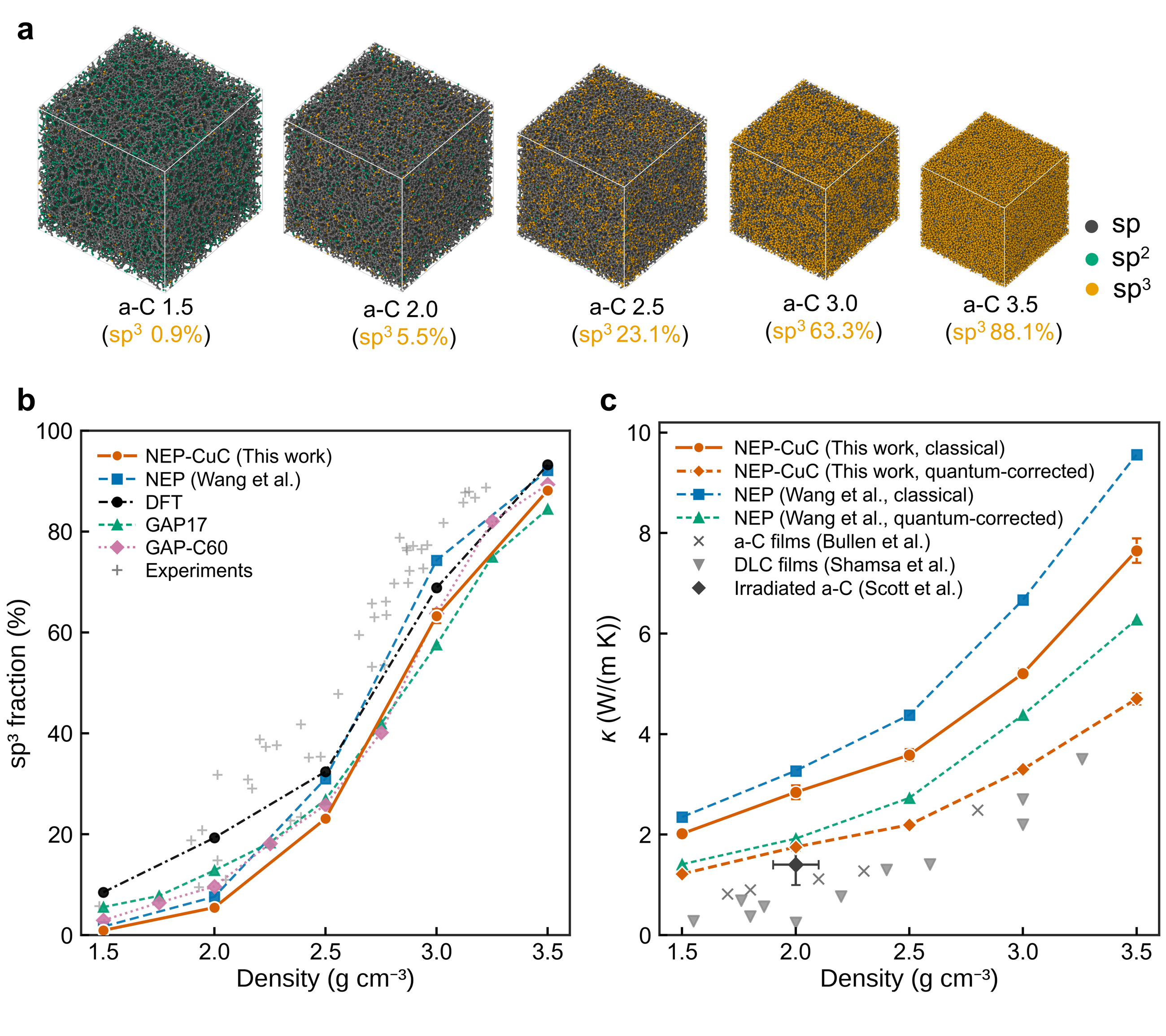}
    \caption{\textbf{Density-dependent coordination and thermal transport in amorphous carbon.} \textbf{a}, Representative melt--quenched structures from NEP-CuC model M1 at densities of 1.5--3.5~g\,cm$^{-3}$, coloured by two-, three- and fourfold C coordination. \textbf{b}, Fraction of fourfold-coordinated C atoms versus density, compared with the NEP model of Wang \textit{et al.}~\cite{PhysRevB.111.094205}, DFT and GAP17~\cite{Deringer2017}, GAP-C60~\cite{Muhli2021} and experimental data~\cite{Fallon1993,Schwan1996,Ferrari2000}. \textbf{c}, Classical and quantum-corrected thermal conductivity at 300~K, compared with the NEP model of Wang \textit{et al.}~\cite{PhysRevB.111.094205} and measurements by Bullen, Shamsa and Scott~\cite{Bullen2000,Shamsa2006,Scott2021}. Panels b,c summarize NEP-CuC results obtained using models M1--M4. In the legends, a-C and DLC denote amorphous carbon and diamond-like carbon, respectively.}
    \label{fig:ac}
\end{figure*}

Using NEP-CuC+D3(BJ), we relax a 40,192-atom commensurate structure with graphene rotated by $10.47^\circ$ relative to the Cu(111) substrate (Methods and Supplementary Note~S4). The relaxed graphene exhibits registry-dependent height modulation above the Cu substrate (\autoref{fig:moire}). Its lateral repeat of 1.346~nm is prescribed by the commensurate cell construction. The morphology is qualitatively consistent with the R10 graphene/Cu(111) superstructures observed by scanning tunnelling microscopy~\cite{Sule2014GrapheneCu}. This calculation provides a qualitative structural check of NEP-CuC+D3(BJ) at a large, weakly bound interface. Graphene-height maps for all five ensemble members, obtained using model-specific equilibrium lattice parameters, are provided in Supplementary Fig.~S9.

\vspace{0.5cm}
\Needspace{5\baselineskip}
\noindent\textbf{Density-dependent structure and thermal transport of amorphous carbon}

\noindent To test four members of the NEP-CuC ensemble on disordered carbon networks, we melt--quenched periodic 125,000-atom cells at densities from $1.5$ to $3.5$~g\,cm$^{-3}$~\cite{Deringer2017,PhysRevB.111.094205}. The mean fourfold-coordinated fraction increases from 0.9\% to 88.1\% (\autoref{fig:ac}a,b). This coordination fraction serves as a structural proxy for $\mathrm{sp}^{3}$ bonding. The bond-angle maximum shifts from $119.5^\circ$ to $107.5^\circ$, consistent with the increasing prevalence of tetrahedral environments (Supplementary Fig.~S10).

The density dependence of fourfold coordination agrees qualitatively with predictions from previous \gls{dft} calculations and carbon-specific \glspl{mlp}~\cite{Deringer2017,Muhli2021,PhysRevB.111.094205}, as well as experimental measurements~\cite{Schwan1996,Fallon1993,Ferrari2000}. Discrepancies are largest at intermediate densities of $2.0$--$2.5$~g\,cm$^{-3}$. At the highest density, the predicted fraction approaches the approximately 92\% value obtained with the carbon-specific NEP potential developed by Wang \textit{et al.}~\cite{PhysRevB.111.094205}.

\begin{figure*}[ht]
    \centering
    \includegraphics[width=1.5\columnwidth]{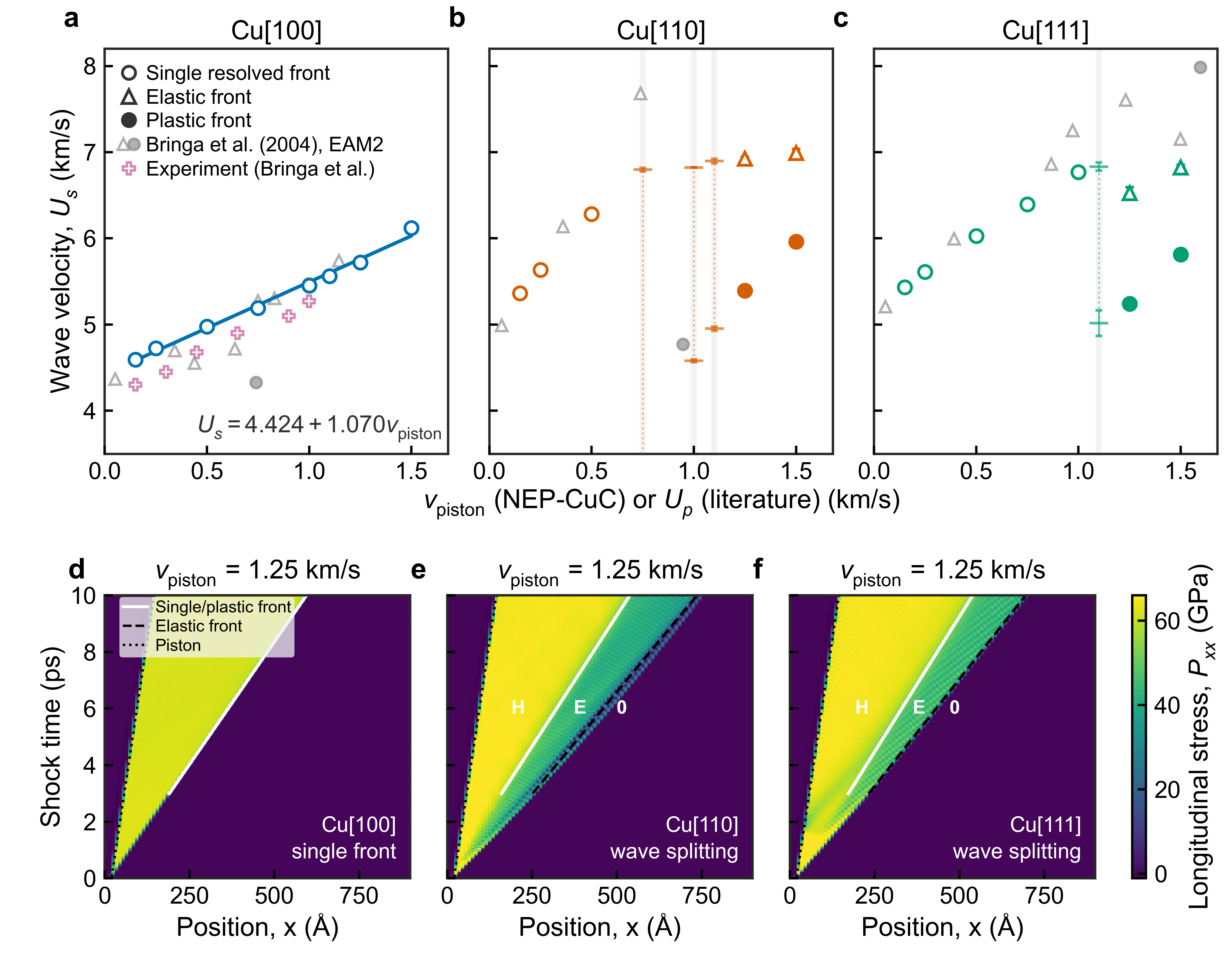}
    \caption{\textbf{Orientation-dependent shock-front splitting in Cu.} \textbf{a--c}, Wave-front velocity along [100], [110] and [111] versus imposed piston velocity, $v_{\mathrm{piston}}$, for NEP-CuC. Literature data retain their reported particle-velocity coordinates, $U_p$. Open circles, triangles and filled circles denote single, elastic and plastic fronts, respectively. Grey bands and horizontal ticks mark transitional cases. Grey markers show EAM2 results (EAM denotes embedded-atom method) and pink crosses show the [100] experimental data, both reproduced from Bringa \textit{et al.}~\cite{Bringa2004CuShock}. The blue line fits the NEP-CuC [100] single-front data. Error bars denote the standard error of the mean (SEM) across NEP-CuC models M1--M3. \textbf{d--f}, Representative longitudinal-stress maps from NEP-CuC model M1 at $v_{\mathrm{piston}}=1.25$~km\,s$^{-1}$. Dotted black, dashed black and solid white lines mark the piston, elastic front and single/plastic front, respectively. Labels 0, E and H denote the initial, elastic-precursor and shocked regions, respectively.}
    \label{fig:cu_single_shock_branches}
\end{figure*}

We then assessed whether these model-generated networks also reproduce the density dependence of thermal transport. At 300~K, the classical conductivity rises from $2.02$ to $7.65$~W\,m$^{-1}$\,K$^{-1}$ as the density increases from $1.5$ to $3.5$~g\,cm$^{-3}$ (\autoref{fig:ac}c). This increase accompanies the coordination changes and is consistent with the previously reported nonlinear density dependence~\cite{PhysRevB.111.094205}. The frequency-dependent harmonic quantum correction (Methods) reduces these values by 36--40\%, to $1.22$--$4.70$~W\,m$^{-1}$\,K$^{-1}$, while preserving the density trend. The corrected conductivities lie closer to the experimental range shown in \autoref{fig:ac}c~\cite{Bullen2000,Shamsa2006,Scott2021}. Together, coordination and transport provide a structure--property test of carbon networks generated by the potential itself.

\vspace{0.5cm}
\noindent\textbf{Transferability under shock compression of Cu}

\begin{figure}[ht]
    \centering
    \includegraphics[width=\columnwidth]{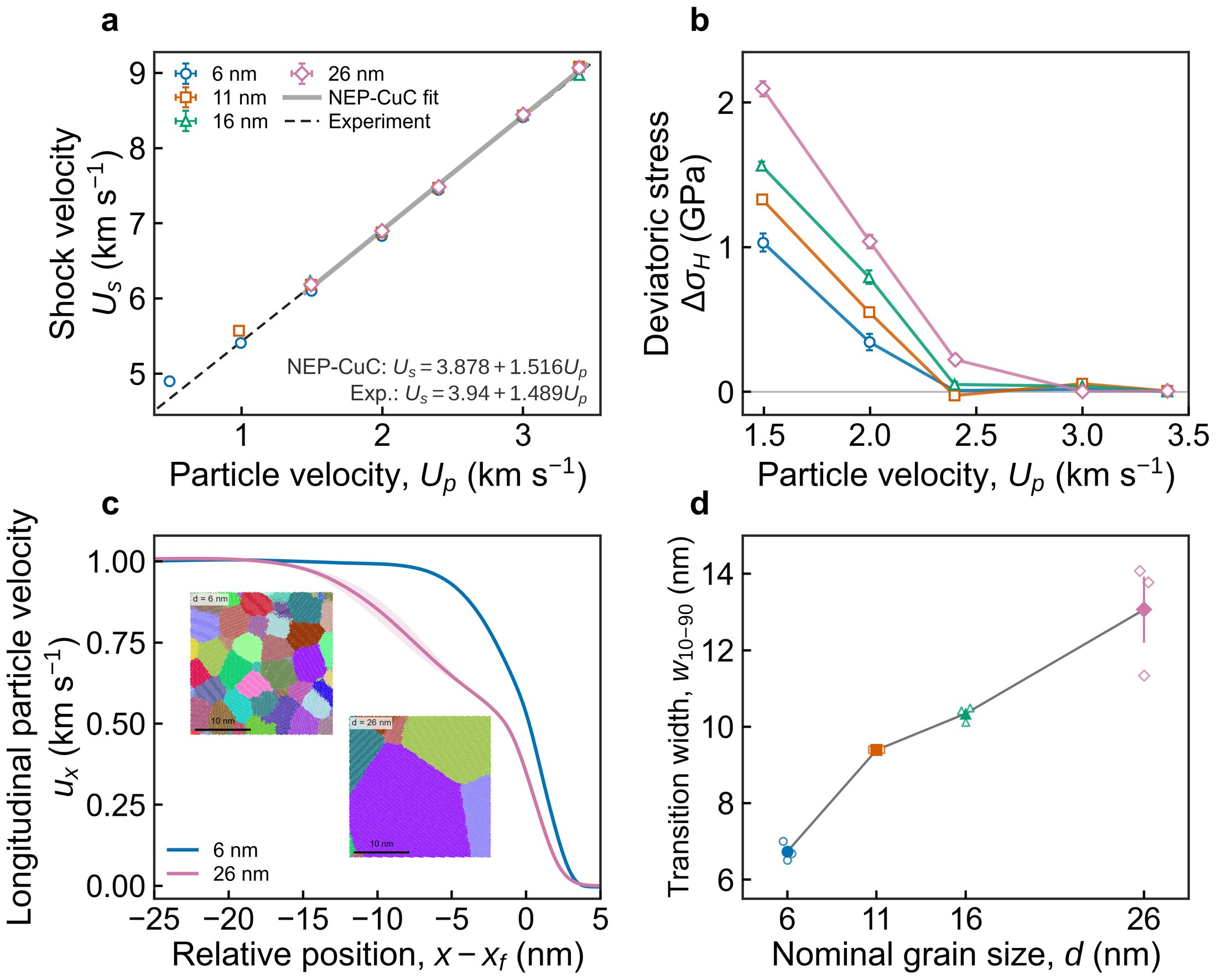}
    \caption{\textbf{Shock response across polycrystalline Cu structures with different grain sizes.} \textbf{a}, Shock velocity, $U_s$, versus measured particle velocity, $U_p$, for nominal grain sizes of 6--26~nm, together with the pooled NEP-CuC fit and the experimental polycrystalline-Cu relation used by Bringa \textit{et al.}~\cite{Bringa2004CuShock}. \textbf{b}, Shocked-state longitudinal--transverse stress difference. \textbf{c}, Leading-front-aligned particle-velocity profiles for the 6- and 26-nm structures at $v_{\mathrm{piston}}=1.0$~km\,s$^{-1}$. Insets show the corresponding grain geometries. \textbf{d}, Composite 10--90\% velocity-transition widths at the same loading velocity. Open symbols show individual models and filled symbols their means. Error bars and bands denote the standard error of the mean (SEM) across NEP-CuC models M1--M3.}
    \label{fig:cu_poly_shock}
\end{figure}

\noindent Amorphous-carbon formation probes bond-network reconstruction, while Cu shock compression tests the metallic response through compression energetics, elastic anisotropy and plasticity. Across the sampled loading range, compression along [100] produces a single resolved compression front, whereas [110] and [111] develop separate elastic and plastic fronts over part of the range (\autoref{fig:cu_single_shock_branches}a--c). At an imposed piston velocity of $v_{\mathrm{piston}}=1.25$~km\,s$^{-1}$, both [110] and [111] show an elastic precursor (\autoref{fig:cu_single_shock_branches}e,f). This orientation dependence is qualitatively consistent with earlier atomistic simulations~\cite{Bringa2004CuShock,Perriot2014CuPrecursor}, although the predicted [100] front velocities exceed experiment over the common loading range (\autoref{fig:cu_single_shock_branches}a). The present wave-front differences are observed over 12~ps in approximately 100-nm-long targets. Simulation protocols and wave-front diagnostics are detailed in Methods and Supplementary Note~S5.

These wave-front differences are accompanied by orientation-dependent defect structures in the shocked region (Supplementary Fig.~S11). At $v_{\mathrm{piston}}=1.25$~km\,s$^{-1}$ and $t=12$~ps, [100] exhibits the largest fraction of atoms classified as hexagonal close-packed (hcp), with multilayer stacking faults dominating its classified planar-fault population. Intrinsic stacking faults account for a larger share of the classified planar faults under [111] loading. The [111] shocked region also has a higher mean dislocation line density than [100], although its hcp-classified fraction is lower.

\begin{figure}[tp]
    \centering
    \includegraphics[width=\columnwidth]{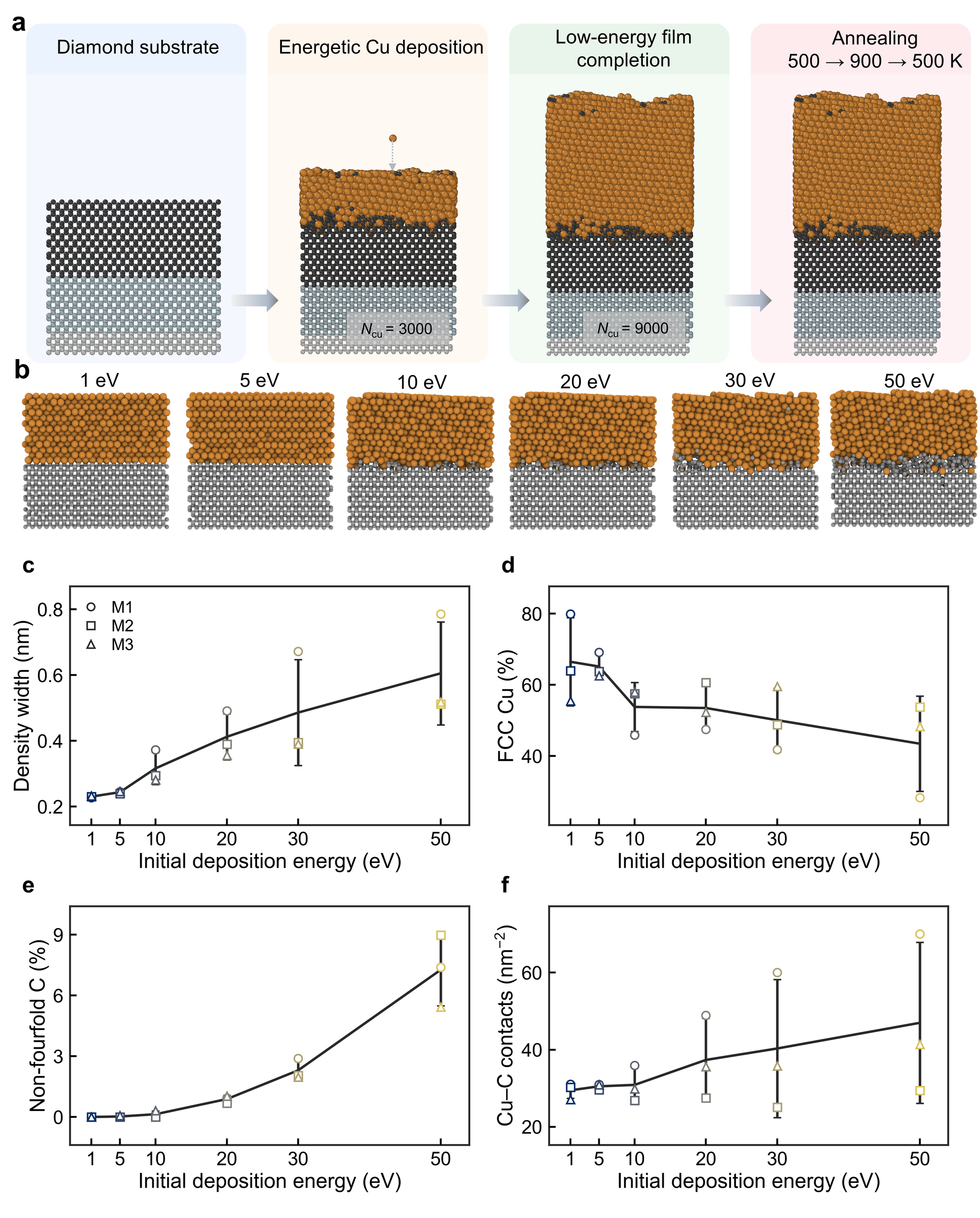}
    \caption{\textbf{Preparation history and the resulting Cu/diamond interphase.} \textbf{a}, Initial Cu deposition to 3,000 retained atoms, growth at 0.3~eV to 9,000 retained Cu atoms, and annealing. $N_{\rm Cu}$ denotes retained Cu atoms. \textbf{b}, Loading-onset configurations from NEP-CuC model M1 for the six initial deposition energies. Cu is orange and C grey. \textbf{c}, Coarse-grained Cu-density 10--90\% transition width, $w_{\rho}$. \textbf{d}, Face-centred cubic (fcc) fraction of Cu within 10~\AA{} above the original diamond surface. \textbf{e}, Non-fourfold C fraction in the layer 2--12~\AA{} below the original surface. \textbf{f}, Total geometric Cu--C contact density, defined as pairs within 2.5~\AA{} per initial in-plane area. Panels c--f are evaluated at the start of tensile loading after constrained re-equilibration. Circles, squares and triangles identify models M1, M2 and M3. Lines and error bars denote means and sample standard deviations (SDs), respectively, across three model--preparation pairs per initial energy.}
    \label{fig:cu_deposition}
\end{figure}

We next extended this assessment to polycrystalline Cu with randomly oriented grains. For $v_{\mathrm{piston}}\geq1.5$~km\,s$^{-1}$, the pooled single-wave states follow $U_s=3.878+1.516U_p$, close to the experimental relation $U_s=3.94+1.489U_p$ used by Bringa \textit{et al.}~\cite{Bringa2004CuShock} (\autoref{fig:cu_poly_shock}a, with velocities in km\,s$^{-1}$). In this loading regime, shock velocities and compressions vary little among the sampled grain structures. Previous simulations of nanocrystalline Cu also reported weak grain-size dependence of Hugoniot pressure~\cite{Sichani2015NanoCuShock}. Independently measured density and longitudinal-stress plateaux are also consistent with the Rankine--Hugoniot mass- and momentum-conservation relations (Supplementary Fig.~S12).

By contrast, more local and transient measures show greater sensitivity to the sampled grain structures. At $v_{\mathrm{piston}}=1.5$~km\,s$^{-1}$, the longitudinal--transverse stress difference increases from $1.03\pm0.06$~GPa for the 6-nm structure to $2.10\pm0.05$~GPa for the 26-nm structure (\autoref{fig:cu_poly_shock}b). In columnar Cu, a nearly grain-size-independent $P$--$V$ Hugoniot was reported alongside grain-size-dependent shear stress~\cite{Hu2021NanoCuHugoniot}. At $v_{\mathrm{piston}}=1.0$~km\,s$^{-1}$, the composite 10--90\% velocity-transition width increases from $6.74\pm0.15$ to $13.06\pm0.87$~nm over the same grain-size range (\autoref{fig:cu_poly_shock}c,d, Supplementary Table~S1). These differences may reflect variations in plastic relaxation and orientation-dependent wave propagation within the sampled microstructures. Overall, the bulk $U_s$--$U_p$ relation is comparatively insensitive to grain structure in the sampled single-wave regime, while stress anisotropy and transient wave-front structure retain a stronger microstructural dependence.

\vspace{0.5cm}
\noindent\textbf{Process-generated Cu/diamond interphases and preparation-dependent mechanical response}

\begin{figure*}[tp]
    \centering
    \includegraphics[width=\textwidth,height=170mm,keepaspectratio]{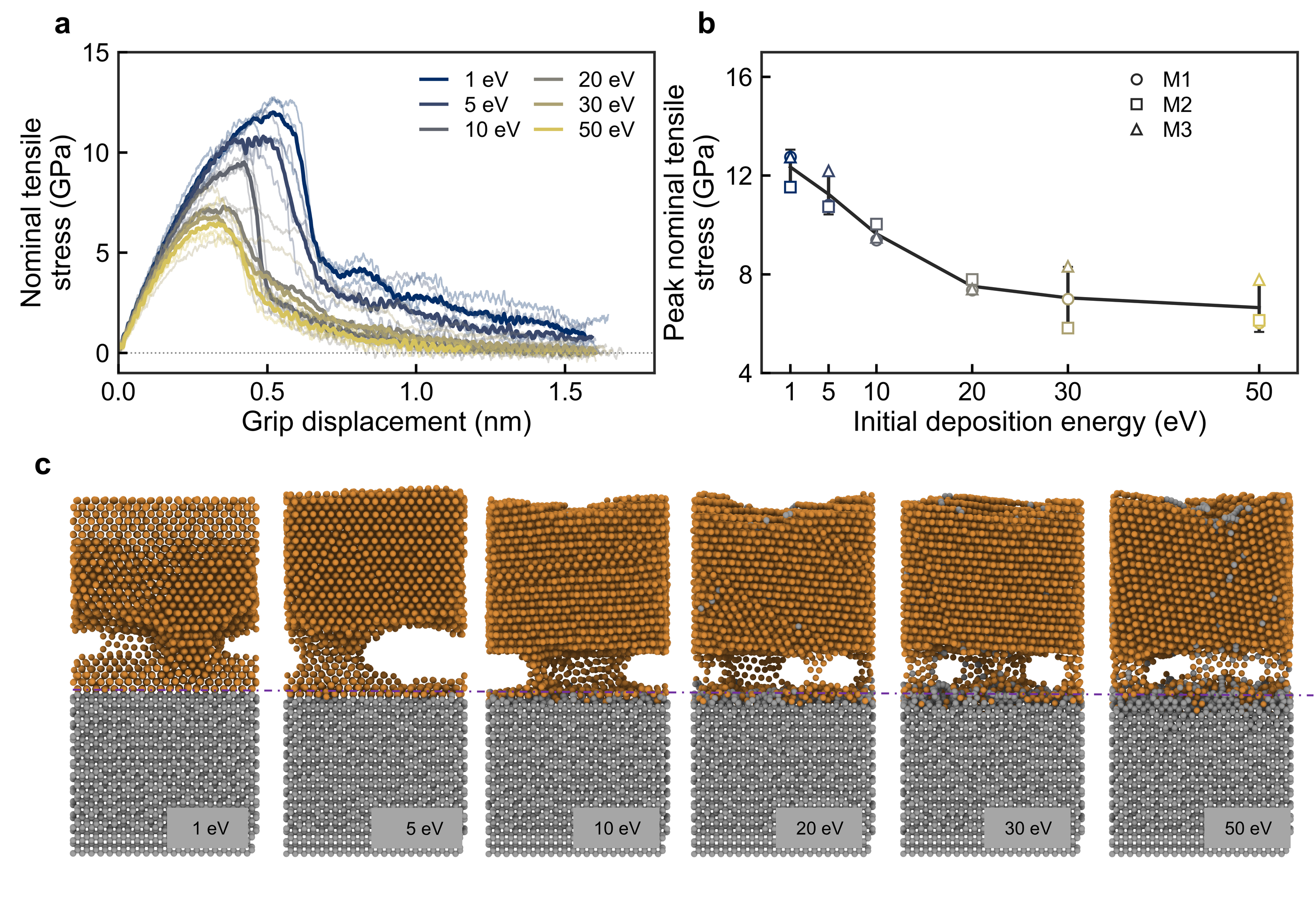}
    \caption{\textbf{Normal mechanical response of deposited Cu/diamond assemblies.} \textbf{a}, Nominal tensile stress versus grip displacement at 500~K and $10^{9}$~s$^{-1}$. Thin curves show individual trajectories. Thick curves show their means over the shared displacement interval. \textbf{b}, Individual peak nominal tensile stresses and means $\pm$ sample SD across three model--preparation pairs per initial energy. Colours identify initial deposition energies, and circles, squares and triangles identify NEP-CuC models M1, M2 and M3. \textbf{c}, Representative post-peak bridging configurations from model M1 for all six initial energies, selected as the first stored frames after nominal tensile stress falls to 20\% of each trajectory's smoothed peak (Supplementary Note~S7). The purple dashed line marks the original diamond surface. Cu is orange and C grey.}
    \label{fig:cu_tensile}
\end{figure*}

\noindent Having tested non-equilibrium response at both elemental limits, we examined their combination during Cu deposition on diamond. Here, an interphase denotes the processing-altered region between deposited Cu and diamond. Energy-selected Cu deposition experiments reveal morphology changes at impact energies of tens of electronvolts~\cite{Pomeroy2000STM}, while measured Cu-ion energy distributions in sputtering vary with discharge conditions~\cite{Atmane2024EnergyDistribution}. We constructed Cu/diamond(001) assemblies by explicit deposition, allowing the interphase to emerge during processing. The controlled initial stage used incident Cu energies of 1--50~eV and continued until 3,000 Cu atoms were retained. Growth at a constant 0.3~eV then completed each film to 9,000 retained Cu atoms (\autoref{fig:cu_deposition}a and Supplementary Movie S1). Representative M1 configurations show more uneven Cu-film surfaces and exposed C atoms at 20 and 50~eV than at 1~eV (Supplementary Fig.~S20). All assemblies underwent the same 500--900--500~K annealing cycle before constrained re-equilibration and tensile loading. Three independently trained NEP-CuC models (M1--M3) were each used throughout the deposition--annealing--loading workflow, enabling comparison of preparation-dependent trends across model--preparation pairs.

\begin{figure*}[tp]
    \centering
    \includegraphics[width=\textwidth]{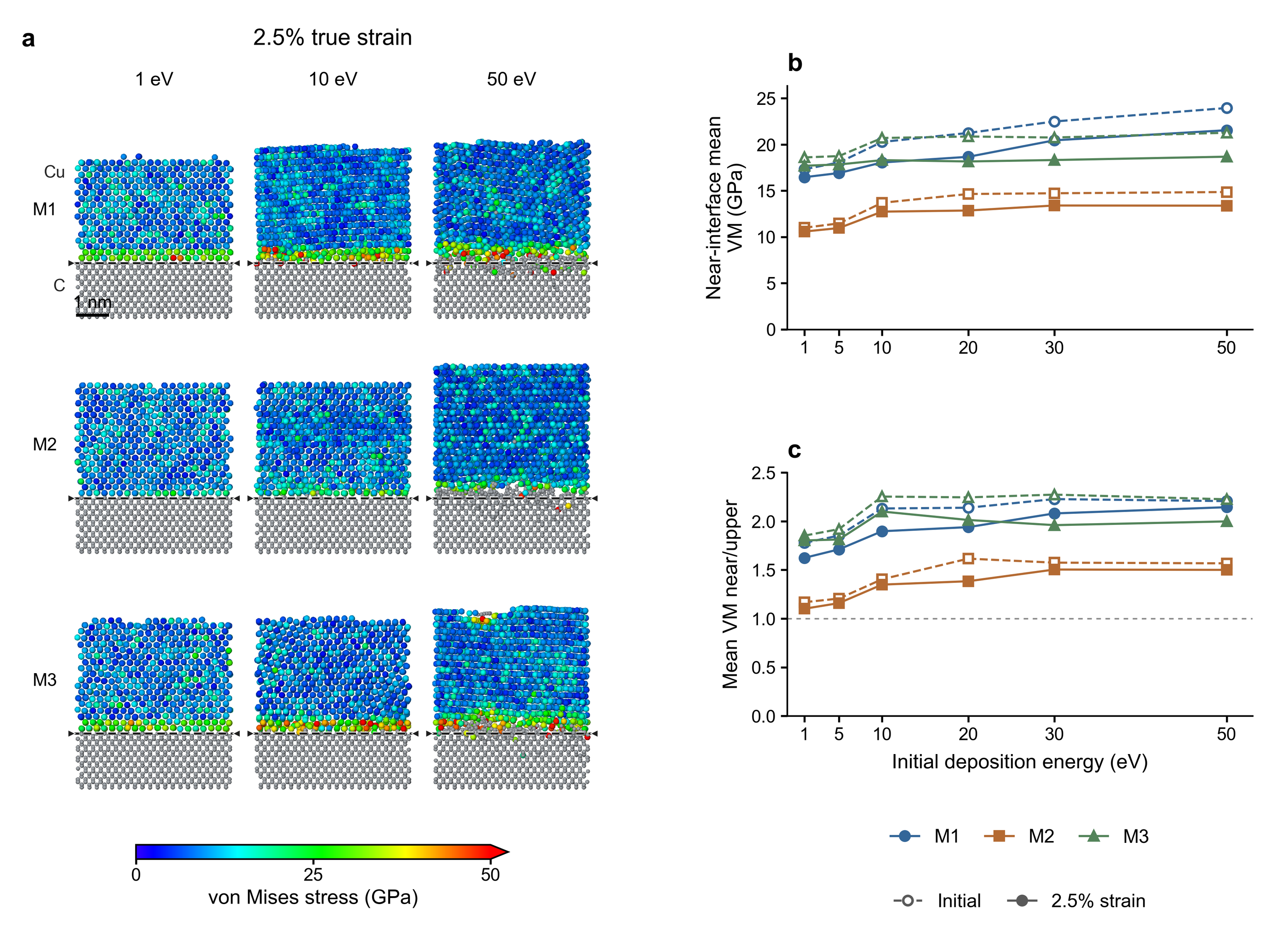}
    \caption{\textbf{Preparation-dependent local stress heterogeneity in Cu/diamond assemblies.} \textbf{a}, Atom-resolved von Mises stress at 2.5\% true strain for initial deposition energies of 1, 10 and 50~eV. Rows identify NEP-CuC models M1--M3. Mobile Cu uses a shared colour scale; C is grey and grip atoms are omitted. Dashed lines mark the original diamond surface. Scale bar, 1~nm. \textbf{b,c}, Near-interface mean von Mises stress and the near-interface/upper-film ratio, respectively, for all six initial deposition energies. Hollow symbols with dashed lines show loading onset; filled symbols with solid lines show 2.5\% strain. Regional means follow fixed initial mobile Cu atoms and include full lateral populations (Methods).}
    \label{fig:cu_local_response}
\end{figure*}

After film completion and annealing, assemblies prepared at higher initial energies retain broader Cu-density transition regions and less ordered near-interface Cu (\autoref{fig:cu_deposition}b--f). The mean coarse-grained Cu-density transition width rises from $0.230\pm0.003$~nm at 1~eV to $0.605\pm0.157$~nm at 50~eV (\autoref{fig:cu_deposition}c). Over the same range, the mean near-interface fcc Cu fraction falls from approximately 66\% to 43\% (\autoref{fig:cu_deposition}d). The mean non-fourfold C fraction in the sampled subsurface layer reaches 7.27\% at 50~eV (\autoref{fig:cu_deposition}e). The mean geometric Cu--C contact density also increases, from $29.48\pm2.10$ to $46.97\pm20.86$~nm$^{-2}$, albeit with substantial variation at high energies (\autoref{fig:cu_deposition}f). Thus, the initial energetic stage leaves a persistent structural signature across Cu order, carbon coordination and interphase geometry. Definitions and sensitivity checks are given in Methods, Supplementary Note~S6 and Supplementary Figs.~S13 and S14.

We then subjected the process-generated assemblies to normal tensile loading to examine the mechanical response of their preparation-dependent structures (Methods and Supplementary Note~S7). At 500~K, the mean peak nominal tensile stress decreases from $12.35\pm0.71$~GPa at 1~eV to $6.65\pm0.98$~GPa at 50~eV (\autoref{fig:cu_tensile}a,b). The mean decline is steeper between 1 and 20~eV than between 20 and 50~eV. Low-energy assemblies sustain higher nominal tensile stresses over a broader grip-displacement range, whereas higher-energy assemblies show lower peak and post-peak nominal tensile stresses. This decline in load-bearing capacity occurs despite increasing mean geometric Cu--C contact density, indicating that contact count alone does not explain the response.

Representative post-peak configurations from model M1 show corresponding differences in deformation morphology (\autoref{fig:cu_tensile}c). At low initial deposition energies, extended Cu ligaments and pronounced necking bridge the film and substrate, consistent with substantial deformation within the Cu-rich region. At higher energies, the remaining Cu bridges are shorter and lie closer to the heterogeneous interphase. Cu bridges remain visible at all six energies in these snapshots, while their changing morphology suggests a progression from extended Cu necking towards more interphase-proximal deformation and separation.

The loading-onset structural trends help interpret these differences in nominal tensile stress and deformation morphology. Paired trajectories show that similar geometric Cu--C contact densities can accompany different peak nominal tensile stresses (Supplementary Fig.~S15). Between 1 and 10~eV, the mean peak nominal tensile stress and mean near-interface fcc Cu fraction both decrease. Over this initial-energy interval, the mean non-fourfold C fraction remains below 0.14\% in the sampled 2--12-\AA{} subsurface layer. At higher energies, greater carbon coordination disorder accompanies interphase broadening and the altered bridging morphology. These covarying descriptors provide a structural context for the nominal tensile stress reduction, while their individual mechanical contributions remain unresolved.

To characterize local stress heterogeneity before peak loading, we compared von Mises stresses at loading onset and a common pre-peak true strain of 2.5\% (\autoref{fig:cu_local_response}). Near-interface Cu has a higher mean von Mises stress than upper-film Cu in all 18 model--energy conditions at both states. At 2.5\% strain, the near-interface/upper-film mean stress ratio ranges from 1.10--1.80 at 1~eV to 1.50--2.14 at 50~eV (\autoref{fig:cu_local_response}c). The contrast is already present at loading onset, and the near-interface mean is lower at 2.5\% true strain than at loading onset in all conditions (\autoref{fig:cu_local_response}b, Supplementary Figs.~S16 and S17). Spatial stress metrics and height profiles are provided in Supplementary Figs.~S18 and S19. These fields characterize preparation-dependent stress distributions without identifying a unique fracture mechanism. Together, the simulations connect deposition history to persistent differences in interphase structure, peak load-bearing response and pre-peak local stress distributions under the specified loading protocol.

\vspace{0.5cm}

\Needspace{7\baselineskip}
\noindent{\textbf{DISCUSSION}}

\noindent NEP-CuC provides a common atomistic framework for linking Cu--C processing, structure and physical response. Previous machine-learned potentials have enabled carbon growth on Cu and thermal-transport studies of Cu/diamond interfaces~\cite{liu_generalized_2025,Sun2025CuDiamond}. Building on these studies, we assess the same trained potential across equilibrium properties, collective dynamics and strongly non-equilibrium processes. Amorphous-carbon formation and Cu shock compression test complementary elemental limits, while deposition on diamond combines carbon-network modification, Cu rearrangement and heterogeneous interactions. This progression provides a physical basis for assessing the potential's use in simulations that follow processing through to subsequent response.

The benchmarks show that reproducible physical trends and accurate absolute predictions require separate assessment. Carbon coordination and conductivity increase with density, and the polycrystalline Cu Hugoniot approaches the experimental relation despite microstructure-sensitive transient wave profiles. Similarly, all five models preserve the phonon-conductance ordering of the two coherent Cu/diamond orientations, while absolute conductance varies appreciably. Underestimated diamond conductivity and inaccurate carbon point-defect energetics further show that quantitative transferability remains property dependent. These findings reinforce the need for physical validation beyond average prediction errors~\cite{Liu2023MLIPMetrics}. Ensemble agreement supports trend consistency within the tested conditions, but does not provide a calibrated prediction-error bound.

The deposition results connect preparation history with interface response in a context accessible to energetic growth experiments. In energy-selected Cu/Cu(111) deposition, STM reveals smoother films at 20~eV and more vacancy islands and increasing roughness at higher energies~\cite{Pomeroy2000STM}. Ar bombardment of diamond can likewise generate nanoscale amorphous-carbon layers before Cu deposition~\cite{Yang2021CuDiamond}. These observations provide experimental context for the present simulations, with differences in substrate, incident species and processing conditions. Here, interphase broadening and reduced Cu order persist after film completion and annealing, while increasing mean geometric Cu--C contact density accompanies lower mean peak nominal tensile stress. At low initial energies, this reduction occurs while carbon in the sampled 2--12-\AA{} subsurface layer remains predominantly fourfold coordinated. The near-interface stress contrast is already present at loading onset, showing that tensile loading begins from a non-uniform stress state. Carbon-modified Cu/diamond interfaces can also show improved scratch-test adhesion~\cite{Zhang2023AmorphousCuDiamond}, highlighting the dependence of the response on interphase structure and the measurement used. Cu ordering, carbon coordination and interphase geometry covary, leaving their separate contributions to the simulated weakening unresolved; the local stress fields do not identify a unique fracture mechanism.

These preparation-dependent responses remain bounded by the interaction model and simulation protocol. Short-range Cu--C repulsion deviates from the sampled DFT reference, and one preparation trajectory per model--energy pair leaves model and preparation variability inseparable. Fixed retained atom counts also allow incident dose and growth time to vary across energies. The nominal tensile stresses characterize finite assemblies at 500~K and $10^{9}$~s$^{-1}$, limiting extrapolation to quasistatic interfacial cohesion. Independent preparations would quantify within-model variability, while targeted DFT checks of compressed Cu--C contacts would assess the short-range impact energetics. Qiu \textit{et al.} used DFT separation energetics to compare the cohesion of competing fracture planes in diamond/cubic boron nitride heterostructures~\cite{Qiu2026DiamondcBNFracture}. Comparable calculations for Cu/diamond could clarify how preparation changes local cohesion. Heat-transport calculations on these same deposited structures would extend the processing--structure--response comparison to their phonon-mediated thermal behaviour. Within these boundaries, NEP-CuC connects bulk phases, heterogeneous interfaces and process-generated Cu--C structures at scales beyond routine first-principles molecular dynamics. This workflow could guide extensions to other metal--carbon systems, combining new metal--carbon reference data with the multielement modelling strategy of UNEP-v1, which describes 16 metals and their alloys~\cite{Song2024UNEPv1}. Such models could support studies of diamond-based thermal management, the effects of dilute carbon on alloy toughness, and carbide formation and interfacial evolution.

\vspace{0.5cm}
\noindent{\textbf{METHODS}}

\noindent{\textbf{Dataset construction and evaluation design}}

\noindent Published Cu, C and Cu--C geometries~\cite{Sun2025CuDiamond,yang_tuning_2026,Song2024UNEPv1,kang_intelligent_2026,liu_enhanced_2026,Fan_2024} and purpose-built structures formed a 15,979-geometry candidate inventory. Added structures covered strain, interface separation and registry, defects, atomic mixing, C-in-Cu environments, non-equilibrium disorder and short-range encounters. Selected configurations were relabelled with the DFT protocol below; original source-library labels were excluded from fitting.

Configuration vectors concatenated the means and standard deviations (SDs) of atomic NEP89 descriptors~\cite{Liang2026NEP89} and were standardized. Structural-family stratification and farthest-point sampling in the full descriptor space~\cite{Qi2024DIRECT,gonzalez1985fps,Imbalzano2018Selection} selected 3,152 configurations. Geometry-based de-duplication removed 425 repeats, leaving 2,727 configurations. Five geometry-grouped, stratified training/validation splits used an approximately 95:5 ratio and seeds 2026--2030, yielding models M1--M5, respectively. The principal-component projection in \autoref{fig:trainingRMSE}a used fitted NEP-CuC descriptors only for visualization.

Evaluation used 10,961 source-derived held-out configurations with energy and force labels, including 8,615 with virial labels. These geometries were excluded from the development set; unselected purpose-built configurations were excluded from evaluation. External tests used 5,417 Cu(111) carbon-growth configurations~\cite{liu_generalized_2025} and 450 nanoporous and amorphous-carbon configurations~\cite{PhysRevB.111.094205}, excluded from all model-development stages. Dataset provenance and evaluation-label conventions are given in Supplementary Note~S1.

\vspace{0.5cm}
\noindent{\textbf{First-principles reference calculations}}

\noindent DFT reference energies, forces and virials used \textsc{vasp}~\cite{kresse1996efficiency,kresse1996efficient}, the PBE functional~\cite{perdew1996generalized} and projector-augmented-wave (PAW) method~\cite{blochl1994projector,kresse1999ultrasoft}. Development labels used a 600-eV plane-wave cutoff, $\Gamma$-centred $k$-point meshes with 0.2~\AA$^{-1}$ spacing and $10^{-6}$-eV electronic convergence, without dispersion corrections. Property benchmarks used the same cutoff and typical $k$-point spacings of 0.15--0.22~\AA$^{-1}$, with system-specific relaxation and smearing settings. Graphite and its diamond phase-energy reference used matched PBE-D3(BJ) calculations. The relabelling input is provided in Supplementary Note~S2.

\vspace{0.5cm}
\noindent{\textbf{Potential training and interaction model}}

\noindent NEP-CuC uses NEP4~\cite{Song2024UNEPv1}, with total energy summed over atomic site energies~\cite{fan2021neuroevolution}. Each site network used a 35-component descriptor, one hidden layer of 80 neurons, and radial and angular cutoffs of 6 and 5~\AA{}, respectively. Five models were trained in \textsc{gpumd}~\cite{zheyong2025GPUMD} using the separable natural evolution strategy and full-batch energy, force and virial fitting.

Universal \gls{zbl} repulsion~\cite{ziegler1985stopping,liu2023large} switched smoothly to zero between 1 and 2~\AA{}. Graphite, its matched diamond phase-energy reference, and graphene/Cu(111) calculations used runtime PBE-D3(BJ)~\cite{grimme2010d3,grimme2011d3bj,Ying2024NEPD3} in every energy and force evaluation. Training inputs and interaction settings are given in Supplementary Note~S3.

\vspace{0.5cm}
\noindent{\textbf{Structural and vibrational benchmarks}}

\noindent Equilibrium properties used independently relaxed structures; single-point comparisons used common DFT geometries with fixed coordinates and cell vectors. Lattice parameters came from energy minimization and elastic constants from small-strain energy fits. Point-defect and C-in-Cu calculations used fixed-cell $3\times3\times3$ conventional supercells, with equilibrium fcc Cu and diamond as elemental references. C solution energies covered octahedral, tetrahedral and substitutional sites. Surface and stacking-fault energies were normalized by exposed and fault areas, respectively. Graphene and nanotube benchmarks evaluated in-plane elasticity, vacancy formation, curvature energies and axial stiffness.

For the fully periodic coherent Cu(100)/diamond(100) superlattice, the independently relaxed work of adhesion was
\begin{equation}
W_{\rm ad}=\frac{E_{\rm C,slab}+E_{\rm Cu,slab}-E_{\rm interface}}{2A}.
\end{equation}
Here, the energies refer to independently relaxed elemental slabs and the superlattice; $A$ is the area of one interface. Reference slabs used matching in-plane vectors, and $2A$ averages the two potentially inequivalent interfaces. Constrained separation curves were referenced to sampled minima and did not determine $W_{\rm ad}$.

Harmonic force constants used \textsc{Phonopy}~\cite{togo2015phonopy} with 0.01-\AA{} displacements at each model's relaxed equilibrium structure. The graphene/Cu(111) benchmark used a 40,192-atom commensurate cell, with graphene rotated by $10.47^\circ$ above eight Cu layers. All atoms were relaxed using the Fast Inertial Relaxation Engine (FIRE) and NEP-CuC+D3(BJ). Cell vectors were fixed, with periodic in-plane boundaries and a nonperiodic surface normal. Cell details, fitting equations and phonon ensemble processing are given in Supplementary Note~S4.

\vspace{0.5cm}
\noindent{\textbf{Thermal transport and amorphous-carbon simulations}}

\noindent The thermal conductivities of bulk Cu and diamond were calculated using NEP-CuC in \textsc{gpumd} at 200, 300, 400 and 500~K with a 1~fs time step. Equilibration comprised 1~ns of zero-pressure Berendsen isothermal--isobaric (NPT) dynamics and 0.5~ns of Nos\'{e}--Hoover-chain canonical (NVT) dynamics. HNEMD ran for 10~ns with $z$-directed fields of $10^{-4}$~\AA$^{-1}$ for Cu and $10^{-5}$~\AA$^{-1}$ for diamond~\cite{liu_enhanced_2026}. Means and SEMs were calculated across five independently trained models, excluding the first 0.2~ns of driven sampling.

Interface NEMD used five models in 130,536-atom (100) and 137,280-atom (111) Cu/diamond cells, approximately $5\times5\times40$~nm$^3$, with fixed ends and $z$-directed heat flow. After 0.1~ns NVT equilibration, Langevin reservoirs at $T\pm T/12$ drove 4~ns production with a 1~fs step. Statistics used the final 2~ns of one trajectory per model, orientation and temperature. Conductance was
\begin{equation}
G=\frac{J}{\Delta T},
\end{equation}
where $J$ is the heat-transfer rate per initial cross-sectional area for one heat-flow path. The temperature jump $\Delta T$ came from profile extrapolation to the central interface. Orientation ratios used model-mean conductances. Calculations included phonon transport but excluded electronic heat flow and explicit electron--phonon coupling.

Amorphous-carbon preparation used M1--M4 in periodic 125,000-atom cubic cells at 1.5, 2.0, 2.5, 3.0 and 3.5~g\,cm$^{-3}$. The Wang \textit{et al.}~\cite{PhysRevB.111.094205} melt--quench schedule used a 0.5~fs step. Cells were held at 9000~K for 100~ps, cooled to 5000~K over 50~ps and held for 50~ps. They were quenched to 1000~K over 0.4~ps and held for 500~ps, then cooled to 300~K over 20~ps. Equilibration and sampling at 300~K each lasted 100~ps. Berendsen thermostats controlled high-temperature and cooling stages; Nos\'{e}--Hoover chains controlled the final 300-K stages. Structural statistics used 40 frames per model at 2.5-ps intervals. A 1.9-\AA{} C--C cutoff~\cite{PhysRevB.111.094205} defined two-, three- and fourfold coordination as proxies for $\mathrm{sp}$, $\mathrm{sp}^{2}$ and $\mathrm{sp}^{3}$ environments, respectively.

Amorphous-carbon HNEMD at 300~K used 0.2~ns equilibration at a 1~fs step, then production at a 0.5~fs step. The driving field was $1.5\times10^{-4}$~\AA$^{-1}$ along $x$. Four-model classical summaries used 6-ns windows at 1.5--3.5~g\,cm$^{-3}$. Spectral heat-current conductivities were multiplied by the harmonic quantum-to-classical heat-capacity factor~\cite{PhysRevB.111.094205},
\begin{equation}
f_{\rm Q}(\nu,T)=\left[\frac{x}{2\sinh(x/2)}\right]^2,\qquad x=\frac{h\nu}{k_{\rm B}T},
\end{equation}
where $\nu$ is frequency, $h$ Planck's constant and $k_{\rm B}$ Boltzmann's constant. 

\vspace{0.5cm}
\noindent{\textbf{Shock compression simulations}}

\noindent Shock simulations used M1--M3 in \textsc{gpumd}. Single-crystal Cu targets measured approximately $100\times18\times18$~nm$^3$, with $(x,y,z)$ bases $([100],[010],[001])$, $([110],[1\bar{1}0],[001])$ or $([111],[1\bar{1}0],[11\bar{2}])$. Each model equilibrated its targets at 300~K through 20-ps NVT, 50-ps zero-pressure NPT and 20-ps NVT stages, using a 1~fs step. A 20-\AA{} wall piston imposed $v_{\mathrm{piston}}=0.15$--1.50~km\,s$^{-1}$ for 12~ps, with a 0.5~fs step and no thermostat.

Polycrystalline targets used three-dimensional Voronoi tessellation~\cite{Hirel2015Atomsk}, random grain orientations and nominal grain sizes of 6, 11, 16 and 26~nm (Supplementary Table~S1). Models shared the initial grain geometry and velocity seed at each grain size. After atomic relaxation, equilibration at 300~K used 10-ps NVT, 30-ps zero-pressure NPT and 10-ps NVT stages, with a 1~fs step. Loading velocities were 0.5, 1.0, 1.5, 2.0, 2.4, 3.0 and 3.4~km\,s$^{-1}$. Piston thickness, loading time and time step matched the single-crystal protocol.

Wave fronts were identified from density, temperature, stress and particle-velocity profiles; velocities came from linear position--time fits. Resolved elastic and plastic fronts were tracked separately; unresolved transitional velocities were retained only as diagnostics. The symbols $v_{\mathrm{piston}}$, $U_p$ and $U_s$ denote imposed piston, measured shocked-state particle and shock-front velocities, respectively.

Polycrystalline shocked-state properties used single-wave plateaux excluding piston- and front-adjacent regions. The longitudinal--transverse stress difference was
\begin{equation}
\Delta\sigma_H=P_{xx,H}-P_{T,H},\qquad P_{T,H}=\frac{P_{yy,H}+P_{zz,H}}{2}.
\end{equation}
Mass and momentum conservation were assessed using the Rankine--Hugoniot relations
\begin{equation}
\frac{\rho_H}{\rho_0}=\frac{U_s}{U_s-U_p},\qquad
P_{xx,H}-P_{xx,0}=\rho_0U_sU_p.
\end{equation}
Subscripts $0$ and $H$ denote initial and shocked states; E denotes the elastic-precursor region.

Composite 10--90\% transition widths used leading-front-aligned velocity profiles at $v_{\mathrm{piston}}=1.0$~km\,s$^{-1}$. \textsc{Ovito}~\cite{stukowski2009visualization} analyses used polyhedral template matching (PTM)~\cite{larsen2016ptm}, fcc planar-fault classification~\cite{ovitoFCCPlanarFaults} and the dislocation extraction algorithm (DXA)~\cite{stukowski2012dxa}. Single-crystal defects were analysed at $v_{\mathrm{piston}}=1.25$~km\,s$^{-1}$ between the piston and single/plastic front. Dislocation density was line length per analysed volume. Analysis windows, normalizations and sensitivity checks are given in Supplementary Note~S5.

\vspace{0.5cm}
\noindent{\textbf{Deposition, tensile loading and local analysis}}

\noindent\textit{Preparation and loading.} M1--M3 each supplied one deposition--annealing--loading trajectory per initial energy, using the same model throughout. Diamond(001) substrates contained 20,000 atoms and measured approximately 44~\AA{} in thickness and 50.6--50.7~\AA{} in-plane. Boundaries were periodic in-plane and nonperiodic along the surface normal. During deposition, the bottom 6~\AA{} was fixed and two adjacent 8-\AA{} substrate layers were thermostatted at 500~K.

Initial deposition used five sequential single-atom impacts per cycle, normal to the surface, at 1, 5, 10, 20, 30 or 50~eV. At 3,000 retained Cu atoms, the incident energy was switched directly to 0.3~eV and maintained until 9,000 Cu atoms were retained. NVE impacts used a 0.1~fs step, followed by at least 3~ps of local Langevin cooling at a 0.25~fs step. Injection patterns, impact durations and position-selection rules are specified in Supplementary Note~S6. Retained-atom endpoints allow incident dose and growth time to differ among preparations.

Films were annealed with Nos\'{e}--Hoover chains at a 0.25~fs step, with the bottom group fixed. Heating from 500 to 900~K over 100~ps preceded a 300-ps hold; cooling to 500~K over 100~ps preceded a 100-ps hold.

Annealed assemblies were re-equilibrated in \textsc{lammps} for 10~ps at 500~K, with the bottom and top 10~\AA{} fixed as grips. Loading retained the preparation model and boundary conditions, fixed the lower grip and translated the upper grip rigidly along the surface normal. Dynamics used a 0.5~fs step and a Nos\'{e}--Hoover thermostat with 0.05~ps damping, excluding the surface-normal velocity component. The grip-centre separation followed $L(t)=L_0\exp(\dot{\varepsilon}t)$ at $\dot{\varepsilon}=10^{9}$~s$^{-1}$, where $L_0$ is its initial value. True strain was $\varepsilon=\ln[L(t)/L_0]$.

\textit{Nominal tensile stress and interphase metrics.} Grip displacement was $\Delta_{\mathrm{grip}}=L(t)-L_0$, and nominal tensile stress was
\begin{equation}
\sigma_n=-\frac{F_{z,\mathrm{upper}}}{A_0},
\end{equation}
where $F_{z,\mathrm{upper}}$ is the upper-grip reaction force and $A_0$ the initial in-plane area; tension is positive. Peak statistics used individual trajectory maxima after centred 1-ps moving-average smoothing. Smoothing sensitivity and post-peak frame selection are given in Supplementary Note~S7.

Loading-onset metrics used heights relative to the original diamond surface. PTM classified Cu within 10~\AA{} above it. Geometric Cu--C contact density counted pairs within 2.5~\AA{} over all C centres, divided by $A_0$. Non-fourfold C fractions used atoms 2--12~\AA{} below the surface with a 1.90-\AA{} C--C cutoff. The density-transition width $w_\rho=z_{90}-z_{10}$ used first ascending 10\% and 90\% crossings of each normalized, smoothed Cu-density profile. Profile binning, normalization and sensitivity checks are given in Supplementary Note~S6.

\textit{Local stress analysis.} Von Mises stresses used symmetrized NEP per-atom configurational stress--volume tensors, excluding kinetic terms and divided by fixed species reference volumes. Loading onset was compared with a common pre-peak true strain of 2.5\%. Regional means followed fixed initial mobile Cu atom IDs across full lateral populations. Near-interface Cu initially lay from $-4$ to 10~\AA{} relative to the original surface. Upper-film Cu extended from 10~\AA{} to 4~\AA{} below the upper grip. Maps showed central 10-\AA{}-thick slices. Stress formulas and spatial statistics are given in Supplementary Note~S7.

\vspace{0.5cm}
\noindent{\textbf{Computational throughput benchmarks}}

\noindent MD throughput used one NVIDIA GeForce RTX 4090D GPU and periodic fcc Cu cells with shared initial positions and velocities at 300~K. NVE dynamics used a 1~fs step. NEP-CuC M1 and NEP89 used native \textsc{gpumd}; MACE-MPA-0, CHGNet and M3GNet used GPU-backed \textsc{ase} calculators with velocity-Verlet integration. Throughput was atom-steps per elapsed production time, reported as the median of three separately launched timings, including integration and software overheads. Versions, warm-up, production lengths and timing criteria are given in Supplementary Note~S8.

\vspace{0.5cm}
\noindent{\textbf{DATA AVAILABILITY}}

\noindent The development dataset and trained NEP-CuC potential files supporting this study are publicly available in the NEP-CuC GitHub repository (\url{https://github.com/zhoufy20/NEP-CuC}). Additional data supporting the findings of this study are available from the corresponding author upon reasonable request.

\vspace{0.5cm}
\noindent{\textbf{CODE AVAILABILITY}}

\noindent The simulations and analyses in this study were performed using publicly available software packages, including \textsc{gpumd}~\cite{zheyong2025GPUMD}, \textsc{lammps}~\cite{thompson2022lammps}, Phonopy~\cite{Togo2023PhonopyImplementation,Togo2023PhonopyReview}, the NEP-LAMMPS interface~\cite{zheyong2025GPUMD,thompson2022lammps}, PyNEP~\cite{PyNEP2024} and GPUMDkit~\cite{Yan2026GPUMDkit}. Custom scripts used for data preparation, post-processing and analysis are available from the corresponding author upon reasonable request.

\vspace{0.5cm}
\noindent{\textbf{\large Author contributions}}

\noindent P.Y. and S.Y. conceived and initiated the study. F.Z. and W.Q. contributed equally to this work and jointly prepared the structures and datasets. F.Z. led the DFT calculations and development of the NEP-CuC model, with contributions from J.L. to hyperparameter testing and model training. W.Q. performed the descriptor-space analysis. F.Z. and Y.C. evaluated the fundamental material properties predicted by NEP-CuC. S.Y., Y.L. and P.Y. supervised the project. All authors discussed the results and reviewed and edited the manuscript.

\vspace{0.5cm}
\noindent{\textbf{\large Competing interests}}

\noindent The authors declare no competing interests.

\vspace{0.5cm}
\noindent{\textbf{\large Acknowledgements}}

\noindent The authors acknowledge financial support from the National Key Research and Development Program (2024YFA1209801), the National Natural Science Foundation of China (Grant No. 12602174), the National Natural Science Foundation of China (Grant No. 12325204), the National Natural Science Foundation of China (GYKP022), and the Xi'an Jiaotong University Young Talent Program (LX6J0240001). W. Qiu acknowledges postdoctoral fellowships from the Sackler Center for Computational Molecular and Materials Science and the Ratner Center for Single Molecule Science at Tel Aviv University. The authors would like to thank Prof. Wengen Ouyang and Mr. Shilin Deng of Wuhan University for providing the training dataset from their previous work. The authors also acknowledge the Open Source Supercomputing Center of S-A-I for providing computing resources.

\vspace{0.5cm}
\noindent{\textbf{\large Correspondence}} and requests for materials should be addressed to Jiahui Liu, Shengying Yue, and Penghua Ying.

\vspace{0.5cm}
\noindent{\textbf{\large References}}
\bibliographystyle{naturemag}
\bibliography{ref}

\end{document}